\documentclass{aa}  

 \usepackage{graphicx}
 \usepackage{txfonts}
 \usepackage{natbib}
 \bibpunct{(}{)}{;}{a}{}{,} 
 \usepackage{hyperref}

\defcitealias{bessellbrett88}{BB88}
\defcitealias{gahmetal07}{G07}
\defcitealias{gahmetal13}{G13}

 \newcommand{\Msun}{\ensuremath{\mathrm{M}_\odot}}

 \newcommand{\mum}{\mu {\rm m}}
 \def\Ks{\hbox{$K$s}}
 
 \def\Js{\hbox{$J$s}}
 \def\J{\hbox{$J$}}
 \def\H{\hbox{$H$}}
 \newcommand{\umag}{{$^{m}$}}

 \def\jshks{\hbox{$J$s$\!H\!K$s}}
 \def\jhks{\hbox{$J\!H\!K$s}}              

 \newcommand{\Htwo}{{\hbox {\ensuremath{\mathrm{H_2}}}}}

 \def\fm{\hbox{$.\!\!^{\rm m}$}}
\begin{document}
    \title{Rosette Globulettes and Shells in the Infrared
      \thanks{Based on observations done at the European Southern
        Observatory, La Silla, Chile (ESO programmes 084.C-0299 and 088.C-0630).}\fnmsep
       \thanks{Appendix A is only available in electronic form at http://www.aanda.org. Tables 4 and 5 are only available in electronic form at the CDS via anonymous ftp to cdsarc.u-strasbg.fr (130.79.128.5) or via (URL here).}}

   \author{M. M. M\"akel\"a
           \inst{1}
           \and
           L. K. Haikala\inst{2,1}
           \and
           G. F. Gahm\inst{3}
           }

   \institute{Department of Physics, Division of Geophysics and Astronomy,\\
              P.O. Box 64, FI-00014 University of Helsinki, Finland\\
               \email{minja.makela@helsinki.fi}
          \and
              Finnish Centre for Astronomy with ESO (FINCA),
               University of Turku, V\"ais\"al\"antie 20, 21500 Piikki\"o, Finland
          \and
              Stockholm Observatory, AlbaNova University Centre,
               Stockholm University, SE-106 91 Stockholm, Sweden
              }

    \date{}


  \abstract
    {Giant galactic \ion{H}{ii} regions surrounding central young clusters show compressed molecular shells, which have broken up into clumps, filaments, and elephant trunks interacting with UV light from central OB stars. Tiny, dense clumps of sub-solar mass, called globulettes, form in this environment.
}
    { We observe and explore the nature and origin of the infrared emission and extinction in such cool, dusty shell features and globulettes in one such \ion{H}{ii} region, the Rosette Nebula, and search for associated newborn stars.
}
    {We imaged the northwestern quadrant of the Rosette Nebula in the near-infrared (NIR) through wide-band \jhks\ filters and narrow-band \Htwo\ 1--0 S(1) and P$\beta$ plus continuum filters using SOFI at the New Technology Telescope (NTT) at ESO. NIR images were used to study the surface brightness of the globulettes and associated bright rims. NIR \jhks\ photometry was used to create a visual extinction map and to search for objects with NIR excess emission. In addition, archival images from \textit{Spitzer} IRAC and MIPS 24 $\mum$ and \textit{Herschel} PACS observations, covering several bands in the mid-infrared (MIR) and far-infrared (FIR), were used to further analyze the stellar population, to examine the structure of the trunks and other shell structures and to study this Rosette Nebula photon-dominated region in more detail.
}
    {The globulettes and elephant trunks have bright rims in the \Ks\ band, which are unresolved in our images, on the sides facing the central cluster. Analysis of 21 globulettes where surface brightness in the \Htwo\ 1--0 S(1) line at 2.12 $\mum$\ is detected shows that approximately a third of the surface brightness observed in the \Ks\ filter is due to this line: the observed average of the \Htwo/\Ks\ surface brightness is 0.26$\pm$0.02 in the globulettes cores and 0.30$\pm$0.01 in the rims. The estimated \Htwo\ 1--0 S(1) surface brightness of the rims is $\sim$ 3--8$\times$10$^{-8}$ Wm$^{-2}$sr$^{-1} \mum^{-1}$. The ratio of the surface brightnesses support fluorescence instead of shocks as the \Htwo\ excitation mechanism. The globulettes have number densities of $n(\Htwo)\sim10^{-4}$ cm$^{-3}$ or higher. Masses of individual globulettes were estimated and compared to the results from previous optical and radio molecular line surveys. We confirm that the larger globulettes contain very dense cores, that the density is high also farther out from the core, and that their mass is sub-solar. Two NIR protostellar objects were found in an elephant trunk and one in the most massive globulette in our study.
}
    {}


   \titlerunning{Rosette Globulettes and Shells in the Infrared}
    \maketitle

	\authorrunning{M. M. M\"akel\"a}

%

\section{Introduction} \label{sect:introduction}

Massive newborn stars in giant molecular clouds ignite and ionize the surrounding molecular gas through the interaction between stellar winds and radiation. This causes the plasma to expand. The surrounding molecular gas is compressed into a shell, which accelerates outwards and breaks up into a network of filaments, clumps and pillars. The latter, so-called elephant trunks, point toward the central cluster. A photon-dominated region (PDR) forms along the shell \citep[e.g.][]{hollenbachtielens97}. In optical images of such \ion{H}{ii} regions the shell features appear as dark silhouettes against the bright nebular background with bright rims facing the central stars. Star formation may proceed in the shell and trunks.

The present study is devoted to the \object{Rosette Nebula}, a well-studied \ion{H}{ii} region \citep[e.g.][and references therein]{hbsfr_rosette}. The northwestern (NW) part of the Rosette Nebula is rich in filaments and trunks connected to the shell, which was noted early from optical images \citep{minkowski49}. The shell moves away from the central OB stars at velocities of more than 20 km s$^{-1}$ as concluded from surveys of associated radio molecular line emission \citep{schnepsetal80, gahmetal06, dentetal09}. Any star formed in this environment will be ejected into interstellar space.

\citet{herbig74} drew the attention to some teardrop-shaped cloudlets in this region and suggested that they once detached from the shell. \citet{gonzalesalfonsoetal94} first observed the small cloudlets in CO and CS and derived \Htwo\ number densities of $\sim10^{3}-10^{4}$ cm$^{-3}$. A survey, based on H$\alpha$ images, of such objects in the Rosette Nebula was conducted by \citet{gahmetal07} (hereafter called \citetalias{gahmetal07}). The cloudlets are round or slightly elongated and appear as dark patches in H$\alpha$. Larger objects may have bright H$\alpha$ rims on the side facing the central cluster, and some have distinct head-tail or teardrop shaped structures with the tail pointing away from the cluster \citepalias{gahmetal07}.
This type of tiny clouds forms a class of its own, and was given the name {\it globulettes} in \citetalias{gahmetal07} to distinguish them from proplyds and the much larger globules spread in interstellar space.

Some globulettes are connected by thin dusty threads to shell features, but the majority are quite isolated. It was found that the globulette size distribution peaks at $\sim$~2.5~kAU, but that most have radii $<$10~kAU. Masses were derived from extinction measures indicating that most globulettes have masses $<$~13~M$_{Jup}$ (Jupiter masses), which currently is taken to be the domain of planetary-mass objects. Objects more massive than 100~M$_{Jup}$ are rare. Besides the Rosette Nebula, globulettes are seen in various \ion{H}{ii} regions \citepalias[and references therein]{gahmetal07}.

A follow-up study of molecular line emission, mainly in the three lowest rotational transitions of CO, of some larger globulettes listed in \citetalias{gahmetal07} was made by \citet{gahmetal13} (hereafter called \citetalias{gahmetal13}). It was inferred that such globulettes contain denser cores, and that the gas is molecular and dense also close to the globulettes surface. Masses were derived from the gas content and found similar to those derived in \citetalias{gahmetal07} based on the dust content. The possibility that some globulettes may be gravitationally unstable and form brown dwarfs or planetary-mass objects were explored in \citetalias{gahmetal07} and \citetalias{gahmetal13}, which should be consulted for more details about the properties of globulettes.

The FUV radiation from the central OB cluster interacts with the dense \Htwo-rich matter of the shell and forms a PDR. Absorption of the FUV radiation by the gas and dust in this region produces strong far-infrared (FIR) and near-infrared (NIR) atomic and molecular emission lines, polycyclic aromatic hydrocarbon emission features, and thermal IR dust continuum. \citep{hollenbachtielens97}. Shock and fluorescent excited \Htwo\ NIR rovibrational lines have been observed in PDRs. Fluorescent \Htwo\ emission was first detected in the reflection nebula NGC 2023 by \citet{gatleyetal87}. It has been also observed in e.g. other reflection nebulae \citep{martinietal99}, planetary nebulae \citep{dinersteinetal88}, and \ion{H}{ii} regions such as M16 \citep{allenetal99}. Observational evidence that favors fluorescence over shock excitation includes narrow \Htwo\ line widths \citep{burtonetal90}, a \Htwo\ 1--0 S(1)/2--1 S(1) ratio of $\sim$~1.5-2.0 \citep{blackvandishoeck87}, and transitions from higher excited states ($\nu\sim 7-8$) \citep{burtonetal92}.

The investigation in \citetalias{gahmetal13} focused on molecular line observations of 16 selected massive globulettes. Some of the NIR observations discussed in the present paper were included in that context. They were used to justify the proposed physical model for the globulettes and to get an independent estimate of their mass.
The detection of bright rims facing the Rosette Nebula central cluster in the SOFI broad filter images and also in particular in the narrow filter covering the \Htwo\ 1--0 (S1) 2.12 $\mum$\ line was discussed in \citetalias{gahmetal13} and fluorescence was suggested as the \Htwo\ excitation mechanism. 

In the present paper we add another field and include observations made in additional filters to the dataset. We will evaluate these observations in more detail and with information from a larger area, supplemented with \textit{Spitzer} NIR and mid-infrared (MIR) and \textit{Herschel} FIR data. We list the NIR properties of all the individual globulettes in our images and examine the \Htwo\ 2.12 $\mum$ surface brightness in some of these.
We map the large-scale structure and extinction of the dense part of the NW dust shell and trunks in the Rosette Nebula and derive independent masses and densities of globulettes from NIR extinction measures.
We also examine the distribution of \Htwo\ emission in the area and evaluate whether the bright rims are caused by shocks or fluorescence. Finally, we search for star formation taking place in the observed region.

Observations and data reduction are described in Sect. \ref{sect:observations}. The results are presented in Sect. \ref{sect:results} and discussed further in Sect. \ref{sect:discussion} together with results collected with the space-born telescopes. As in \citetalias{gahmetal07}, we will use 1400~pc as the distance to Rosette Nebula.


\section{Observations and data reduction} \label{sect:observations}

\subsection{NIR Imaging} \label{sect:imagingdata}


The Son of Isaac (SOFI) NIR instrument on the New Technology Telescope (NTT) at the La Silla Observatory, Chile was used to observe the NW area of the Rosette Nebula. This area was previously imaged in narrow-band H$\alpha$ at the Nordic Optical Telescope (NOT) for an examination of cool shell structures \citep{carlqvistetal03} and globulettes \citepalias{gahmetal07}. Five SOFI fields were imaged,  and they will be referred according to the H$\alpha$ observations in \citetalias{gahmetal07} as F(ield)7, F8, F14, F15, and F19, going from west to east. First four fields overlap at the edges. Field 14 was not included in \citetalias{gahmetal13}.
Observations were done in on-off mode in broad \J, \H, and \Ks\ filters to conserve the surface brightness of the globulettes, and in jitter mode in the narrow-band NB 2.124 \Htwo\ 1--0 S(1) at 2.12~$\mum$ and the adjacent continuum NB 2.090 $\mum$ filters. The SOFI field of view is 4\farcm9 $\times$ 4\farcm9 with a pixel size of 0\farcs288.
The observations were carried out in Dec. 2009 and Jan. 2010.

Three OFF fields outside the Rosette Nebula were observed for sky estimates. One OFF field was selected as a reference field for the photometry. Jittered \jshks\ observations of fields roughly corresponding to F8 and F19 were done in Feb. 2007, and jittered NB 2.167 Br$\gamma$ observations at 2.17~$\mum$ of F14 in Jan 2010. SOFI jittered images of all five fields were collected in the NB 1.282 P$\beta$ filter at 1.28~$\mum$ in 2012. Field 15 was also observed in the NB 1.257 filter at 1.26~$\mum$ which covers the continuum adjacent to the P$\beta$\ line.

The on-off observations were done in observation blocks (OBs) of a format OFF1-ON1-OFF2-ON2, acquiring 13 frames of one-minute integration (six ten-second subintegrations) of both ON fields in a single OB. 

The jittered observations were done in standard jittering mode with a jitter box width of 30\arcsec\ (40\arcsec\ in \Js/F19). Jittering enhances the brightness gradients in the image but does not affect stars and galaxies. It also smears out any surface brightness features larger than the jitter box width. The total integration time obtained in different filters is listed in Table \ref{table:observations}. The NB 2.090 continuum filter is narrower than the NB 2.124 \Htwo\ 1--0 S(1) filter, and hence longer integration times were used to achieve similar S/N level.

\begin{table*}
 \caption{A summary of the observations. The fields are referred to with their designations. The coordinates are in J2000.0. The columns list the total ON-integration time in the filter in minutes. The final column lists observations in other filters and the total integration time in that filter. The P$\beta$ observations are from 2012, and the jittered \jshks\ observations from 2007.}
 \label{table:observations}
 \centering
 \begin{tabular}{c c c c c c c c c c c c c}
 \hline\hline
 Field & \multicolumn{2}{c}{Coordinates} & \multicolumn{3}{c}{On-off} & \multicolumn{7}{c}{Jitter} \\
       & RA & Dec & \J & \H & \Ks & \Js & \H & \Ks & P$\beta$ & 2.09 & \Htwo\ & Other \\
 \hline
   F7 & 06:31:00.9 & 05:07:03.6 & 26 & 26 & 26 & - & - & - & 20 & 50 & 40 & - \\
   F8 & 06:31:17.9 & 05:08:05.0 & 26 & 26 & 26 & 20 & 20 & 20 & 40 & 50 & 40 & - \\
   F14 & 06:31:35.5 & 05:08:37.6 & 39 & 39 & 39 & - & - & - & 40 & 50 & 40 & Br$\gamma$: 20 \\
   F15 & 06:31:39.8 & 05:11:16.1 & 26 & 26 & 26 & - & - & - & 40 & 60 & 53 & 1.257: 60 \\
   F19 & 06:32:20.8 & 05:14:06.0 & 13 & 13 & 13 & 20 & 20 & 20 & 40 & 50 & 40 & - \\
   OFF & 06:28:25.2 & 05:21:28.7 & 20 & 20 & 20 & - & - & - & - & - & - & - \\

\hline
\end{tabular}
\end{table*}

Standard stars from the faint NIR standard catalog by \citet{perssoncatalog} were observed before and after each broad-band block. The seeing ranged from 0\farcs7 to 1\farcs0 during the observations.



\subsection{Data reduction} \label{sect:datareduction}

For the data reduction, IRAF\footnote{IRAF is distributed by the National Optical Astronomy Observatories, which are operated by the Association of Universities for Research in Astronomy, Inc., under cooperative agreement with the National Science Foundation.} and the external package XDIMSUM were used.
The ESO SOFI website provides flat field, bad pixel, and illumination correction files that were used. Both ON and OFF frames went through crosstalk removal, bad pixel masking, cosmic ray removal, and stellar object masking. Then the OFF sky from four temporally closest frames was removed from the ON frames, the flat fields and illumination corrections were applied, and the ON frames were coadded.
The jittered observations were reduced in the same manner as the on-off observations except for the sky removal step where two neighboring ON fields were used to estimate the sky brightness instead of separate OFF fields.
Zero-points were derived for each broad-band observation block from the standard star observations. 

Jittering in the on-off observations was done after every OFF1-ON1-OFF2-ON2 cycle, but not inside the cycle. This causes negative, star-like artifacts in the co-added image especially in the ON1 frames (which here are F8 and F15). In some \J\ images a wide dark lane runs through the middle of the co-added frame. This is because there are differences in the sky background levels because of air mass changes in the ON and OFF fields due to the large distance between them. This causes an incomplete flat fielding.

Source Extractor v. 2.5.0 \citep{SExtractor} was used to extract the \jhks\ magnitudes and their formal error for all the objects in the observed fields. The detections from each band were matched and a photometry catalog was created. Objects that were detected in all bands and had positive fluxes were listed in the initial photometry catalog. The magnitudes were transformed from SOFI instrumental magnitudes into the Persson photometric system and then into the 2MASS system as described in \citet{ascensoetal07}.

The final catalog was compiled using the steps described in \citet{cgpaper13} with the difference that objects less than 30 pixels from the frame edges and objects with stellar index less than 0.9 in more than one band were removed. 

The consistency of the SOFI photometry in adjacent fields was checked using the stars in the overlapping regions. The differences of the magnitudes for stars brighter than 18\umag\  in the overlapping regions are  -0\fm04$\pm$0\fm007 (rms 0\fm06) in \J, -0\fm01$\pm$0\fm005 (0\fm05) in \H, and -0\fm01$\pm$0\fm008 (0\fm08) in \Ks. No systematic tendencies were detected in any filter.

The separate SOFI catalogs of each field were then combined to make one catalog. The magnitudes of the stars in the overlapping regions were averaged.
The combined ON catalog has 1777 objects and the chosen OFF field catalog has 389 objects. The limiting magnitudes for a formal 0\fm15 error are 21.0 in \J, 20.0 in \H, and 19.5 in \Ks\ except for F19 where they are 0\fm5 brighter and for F14 and OFF where they are 0\fm5 fainter. The maximum and average errors in magnitudes in the ON catalog are, 0.129 and 0.019 in \J, 0.079 and 0.019 in \H, and 0.126 and 0.029 in \Ks, respectively.

In F19 the jittered \jshks\ data was also combined with the on-off data by scaling the on-off frames to the level of the jittered observations. The \jshks\ photometry catalog based on the combined coadded frames increases the limiting magnitudes in F19 by 0\fm5 when compared to the on-off observations, but the area where the jittered and on-off observations overlap is smaller than just the on-off area, leading to a smaller number of objects in the catalog (302 vs. 345 objects) for this field.

\subsection{Archive data} 

\subsubsection{Spitzer} \label{sect:spitzerdata}

\textit{Spitzer} S18.18 pipeline processed mosaic images of the SOFI fields discussed in this paper were retrieved from the \textit{Spitzer} Heritage Archive. They were imaged with the IRAC instrument at 3.6, 4.5, 5.8, and 8.0 $\mum$\ (channels 1, 2, 3, and 4, respectively) and with the MIPS instrument at 24 $\mum$ but the latter does not include F7. The FWHM in the IRAC images is 1.7-2.0\arcsec\ \citep{fazioetal04} and 6\arcsec\ at 24 $\mum$. These resolutions are too low to detect the smallest globulettes.
The 5.8 and 8.0 $\mum$ band images produced by the S18.18 pipeline need to be multiplied with correction factors 0.968 and 0.973, respectively, due to a calibration error\footnote{\texttt{http://irsa.ipac.caltech.edu/data/SPITZER/docs/\allowbreak{}irac/iracinstrumenthandbook/73/}}. 
Aperture photometry was performed for selected targets using the MOPEX\footnote{\texttt{http://irsa.ipac.caltech.edu/data/SPITZER/docs/\allowbreak{}dataanalysistools/tools/mopex/}} software. The ON aperture radius was chosen as 3.6\arcsec, and the OFF flux was measured in an annulus with radii 3.6\arcsec\ and 8.4\arcsec. The fluxes in each band were multiplied by the appropriate aperture corrections and the ON-OFF flux was transformed into magnitudes using the IRAC zeropoint magnitudes. The values for these were taken from the IRAC Instrument Handbook\footnote{\texttt{http://irsa.ipac.caltech.edu/data/SPITZER/docs/\allowbreak{}irac/iracinstrumenthandbook/}}. The mosaics have a typical error of $\sim$10\% in the flux\footnote{\texttt{http://irsa.ipac.caltech.edu/data/SPITZER/docs/\allowbreak{}irac/iracinstrumenthandbook/84/}}.

\subsubsection{Herschel} \label{sect:herscheldata}

\textit{Herschel} observed the center of the Rosette Nebula and its northern part in the parallel PACS/SPIRE mode (PI. S. Molinari, PID: OT1\_smolinar\_5). Because of the pointing separation of the PACS and SPIRE instruments on the sky, the northern part is covered only by the PACS 70 and 160 $\mum$ bands. These PACS pipeline level 2.5 MADmap processed images were retrieved from the \textit{Herschel} archive. 


\section{Results} \label{sect:results}


\subsection{\jhks\ Imaging}

\begin{figure*}
 \resizebox{\hsize}{!}{\includegraphics{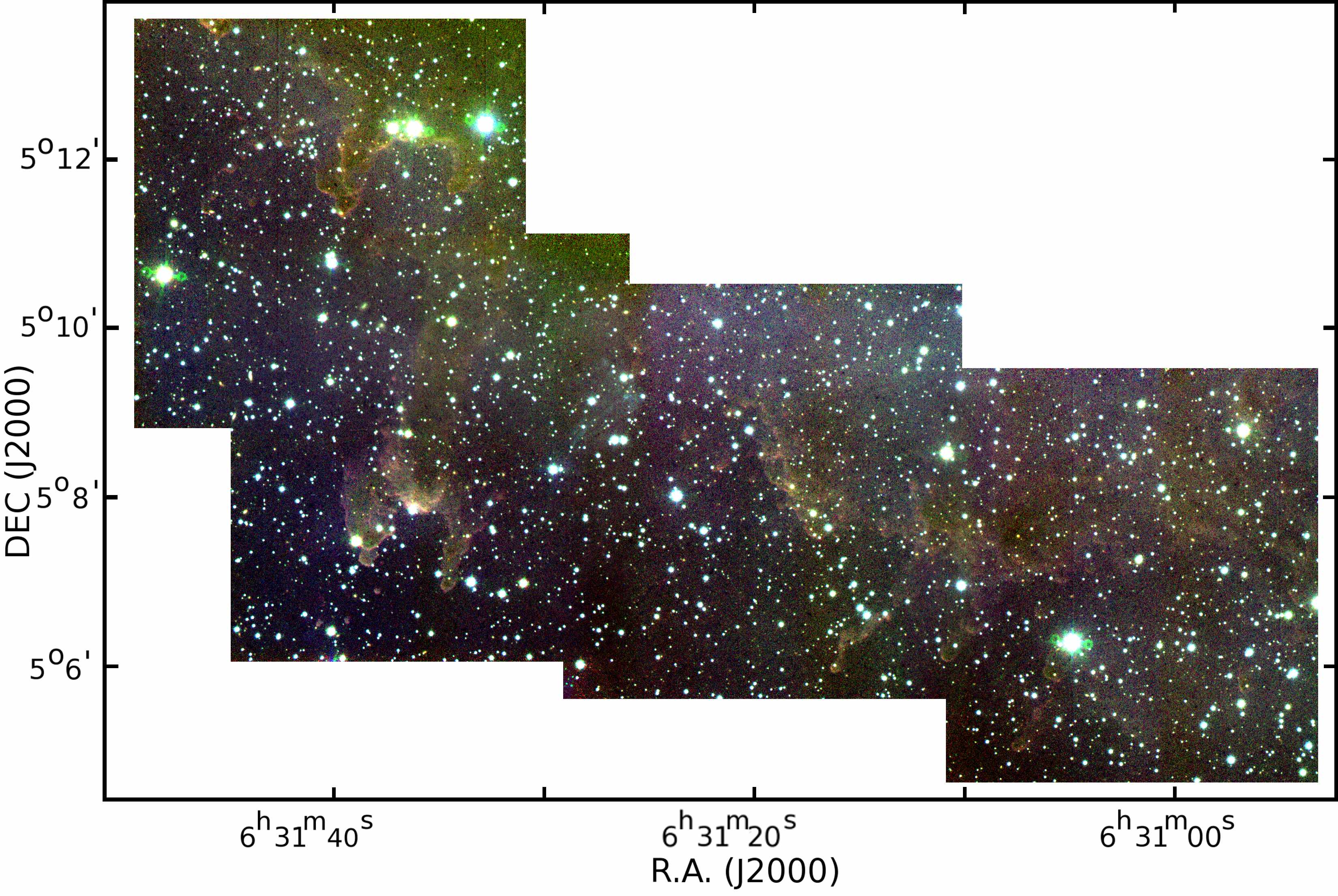}}
    \caption{False-color image of fields 7, 8, 14, and 15 (from west to east). \J\ is coded in blue, \H\ in green, and \Ks\ in red.}
 \label{fig:all4jhk}
 \end{figure*}

\begin{figure*}
 \centering
 \includegraphics [width=17cm] {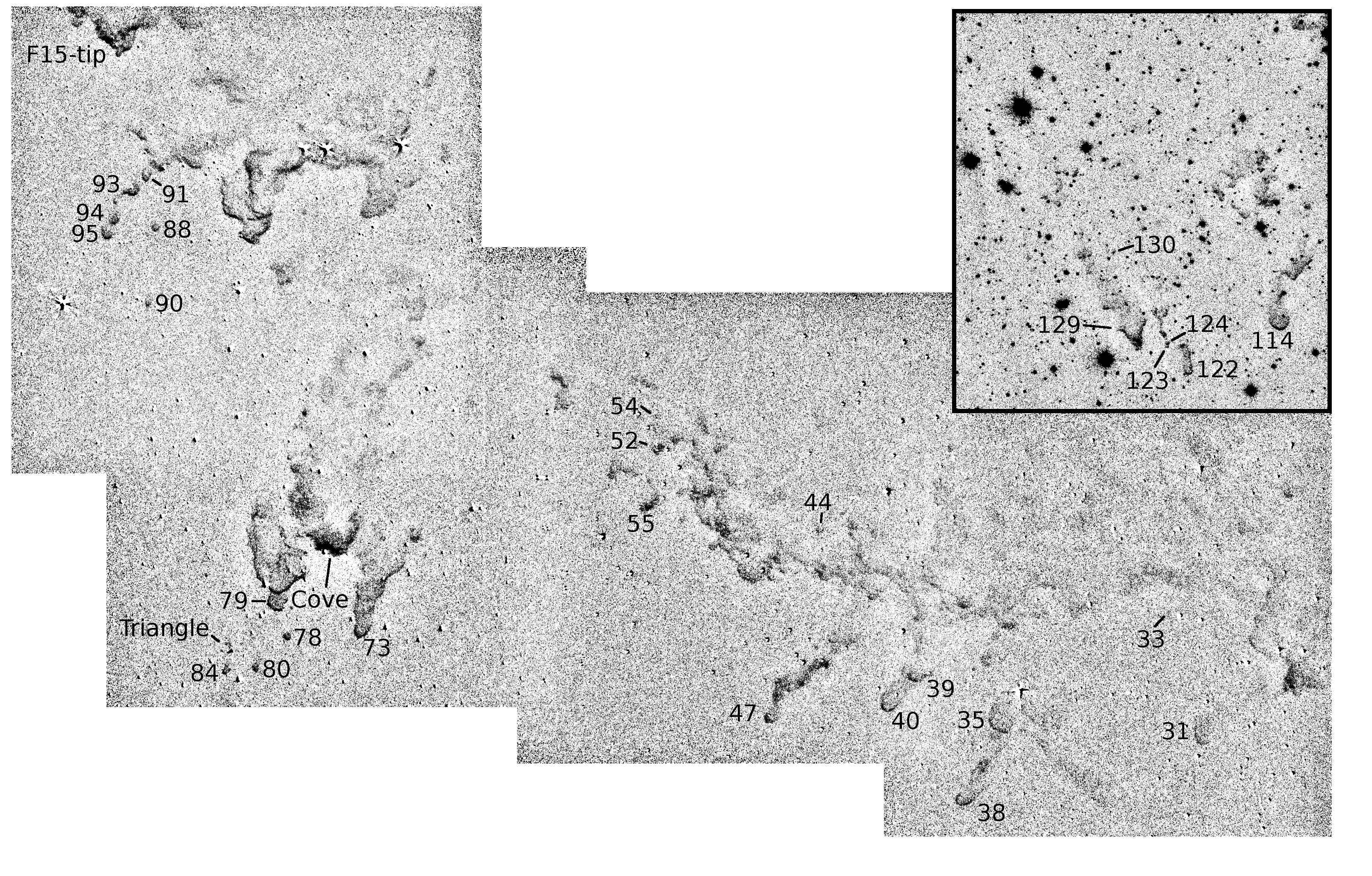}
    \caption{The IDs of the most important globulettes discussed in the paper. The gray-scale image is an enhanced \Htwo\ 2.124 $\mum$ - 2.09 $\mum$ difference image. In the top right-hand corner is the F19 NB 2.124 \Htwo\ S(1) image where the notable F19 globulettes have been marked.}
 \label{fig:globlabels}
 \end{figure*}

A false-color \jhks\ image of fields 7, 8, 14, and 15 is shown in Fig. \ref{fig:all4jhk}. Globulettes with sharp edges facing the center of the Rosette Nebula are seen in all fields. Some globulettes have diffuse tails directed away from the center. Most globulettes show bright rims in \Ks. In the \H\ band the rims are less bright and especially in \J\ the rims are not conspicuous because strong background surface brightness dominates the images.Wispy filaments are seen as line emission between the Wrench and the shell in F8 in the \J\ and \H\ bands.
Enhanced surface brightness extends over the northwestern part of the F7, F8, F14, and F15 images. This surface brightness is seen in the on-off images but not in the jittered images where it is smeared out, and it matches the shell features seen in optical images and CO studies.
The arc-minute-scaled elephant trunks called ``the Wrench'' and ``the Claw'' \citep[nomenclature from][]{gahmetal06} dominate in F14 and F15, (Fig. \ref{fig:all4jhk} east and northeast (NE), respectively). A string of globulettes (hereafter called ``the String'') lies east of the Claw. A group of small globulettes lie south of the Wrench close to the eastern ``jaw''. Fields 7 and 8 cover parts of the larger dust shell. Several globulettes are connected to this shell by thin filaments that resemble small, curved elephant trunks. 
Field 19 is shown in false-color in Figs. \ref{fig:f19jhk} and \ref{fig:f19jhkjitter} in on-off \jhks\ and in jittering \jshks\ modes, respectively. It is located $\sim$10\arcmin\ east of the Claw and contains three large globulettes and some smaller ones but no field-wide shells or trunks. The imaged elephant trunks and shells and also some of the globulettes (e.g. RN 31, 35, 44, 88, 94, 95, 114, 122, 129) are seen in absorption against the bright background emission in \J, indicating that these objects are very dense.
The designations and locations of the globulettes can be found in \citetalias{gahmetal07}. In Fig. \ref{fig:globlabels} we have marked a few objects that are discussed more in detail in the paper. Comparing the \J\ and \Ks\ images with the NOT H$\alpha$ images \citepalias{gahmetal07} shows that the apparent size in most of the globulettes is similar throughout all these wavelengths.

\subsubsection{Jittering vs. on-off}

The on-off observation mode was chosen to detect extended NIR surface brightness in the Rosette Nebula. Contrary to the more commonly used jittering observing mode, the on-off mode preserves such a surface brightness. This can be verified by comparing the jittered F8 \jshks\ image and the on-off measurements shown in Fig. \ref{fig:f8jhkjitter}, left and right panels, respectively.
The surface brightness covering the upper part of the on-off image is absent in the jittered image as structures larger than the jittering width (30\arcsec), such as the continuous dust shell, are smeared out by the jittering process as the background is estimated using the observed field itself instead of a separate off field. However, jittering increases the contrast of the brightness gradients and brings out the bright globulette rims better than the on-off image.
The jittered observing mode is more efficient and the obtained signal-to-noise is better than that of the on-off observations when equal  telescope times are used. These effects combined improve the chance to detect the globulettes and the bright rims. For example the blue-green wisp between the Wrench and the F8 shell in Fig. \ref{fig:all4jhk} is seen better in the jittered images. Even if the differential surface brightness is preserved in the on-off images, the absolute surface brightness level is lost in the data reduction process.

\subsection{NB 2.124 \Htwo\ S(1) and NB 2.090 imaging}

\begin{figure}
 \centering
 \includegraphics [width=8cm]{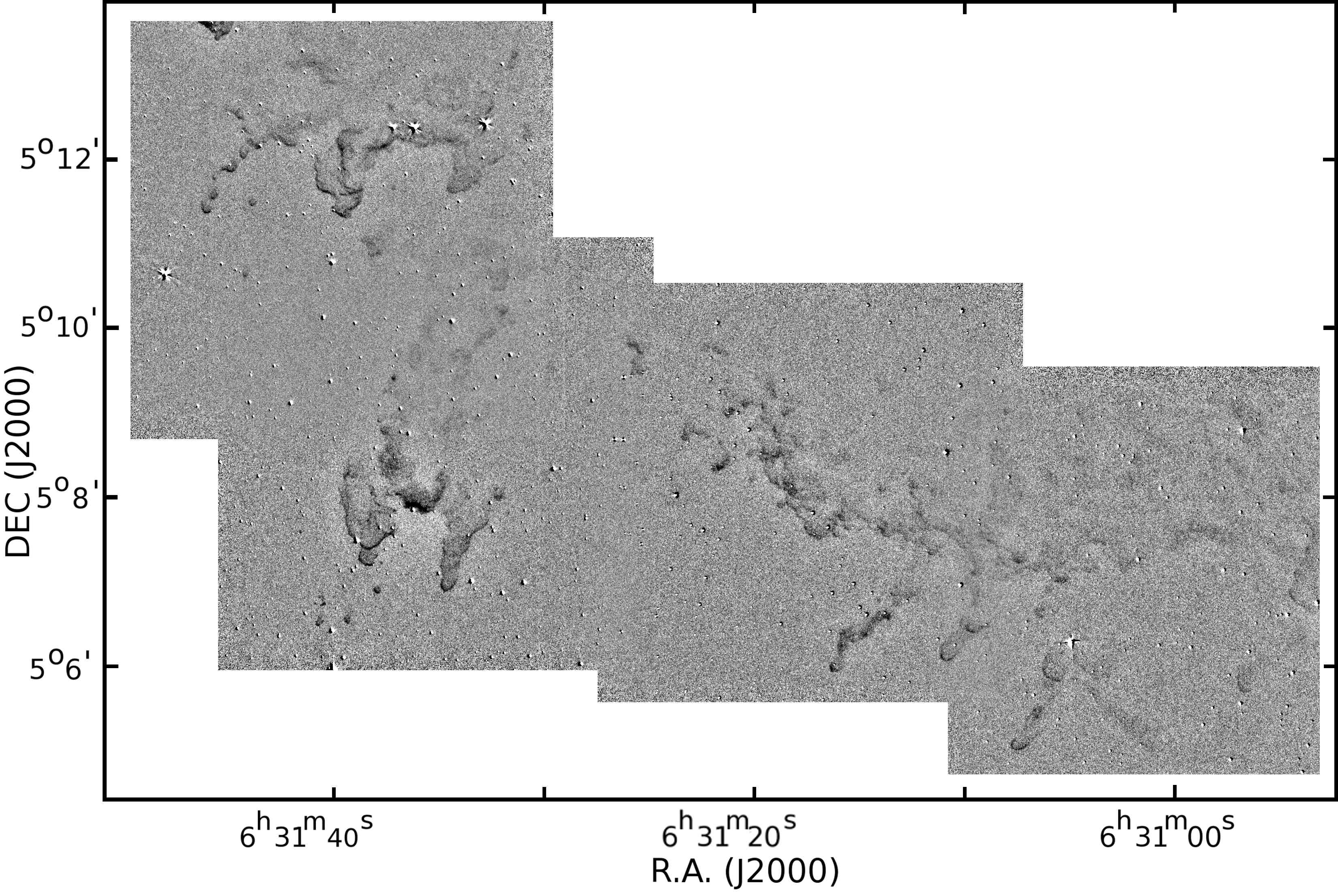}
    \caption{Combined NB 2.124 \Htwo\ S(1) and NB 2.090 difference image (negative) of fields 7, 8, 14, and 15.}
 \label{fig:h2minuscontmap}
 \end{figure}

The NB 2.124 \Htwo\ S(1)--NB 2.090 difference image of the four overlapping fields is shown in Figure \ref{fig:h2minuscontmap}. The original NB 2.124 \Htwo\ S(1) and NB 2.090 images are shown in Figs. \ref{fig:h2map} and \ref{fig:contmap}, respectively. Because the observations are jittered, the gradients are enhanced and the possible extended surface brightness is smeared out.
The features in the NB 2.124 \Htwo\ S(1) image are very similar to those in the \Ks\ band image. In Fig. \ref{fig:h2minuscontmap} the elephant trunks and about 70\% of the globulettes have a bright rim. These rims face the Rosette Nebula central cluster of stars and are unresolved in the images. Some globulettes without a bright rim are seen through their surface brightness.
The stellar diffraction spikes rotate during observations because NTT is an altazimuth telescope and the seeing also varies between the line and continuum observations. These cause an incomplete subtraction of bright stars in the difference images.

Examples of striking rims are in RN 35, 39, 40, 93, 95, and in the Wrench and the Claw. RN 78 and 94 show an \Htwo\ 2.12 $\mum$ halo. The \Htwo\ 2.12 $\mum$ surface brightness in the cores of the smallest globulettes is difficult to separate from the rim brightness in cases where the globulette is so tiny that the bright rim covers the entire globulette (e.g. RN 123, 124, 130).
Table \ref{table:globulettes} lists all the globulettes of \citetalias{gahmetal07} that are located in our NIR frames and their visually determined characteristics. Column 1 lists the globulette number and columns 2 and 3 indicate if the globulette has a bright rim or surface brightness seen in the NB 2.124 \Htwo\ S(1) image. The last two items on the list refer to the eastern jaws of the Wrench and the Claw.

No signs of the globulettes or the elephant trunks are seen in the NB 2.090 continuum image (Fig. \ref{fig:contmap}). This indicates that the observed emission in the NB 2.124 \Htwo\ S(1) image (Fig. \ref{fig:h2map}) is not due to scattered light but comprised of \Htwo\ 1--0 S(1) line emission. The only surface brightness structure seen in the NB 2.090 image is a very faint patch of scattered light around the star in the cove between the jaws of the Wrench (hereafter called ``the Cove'').

\begin{figure}
 \centering
 \includegraphics [width=8cm]{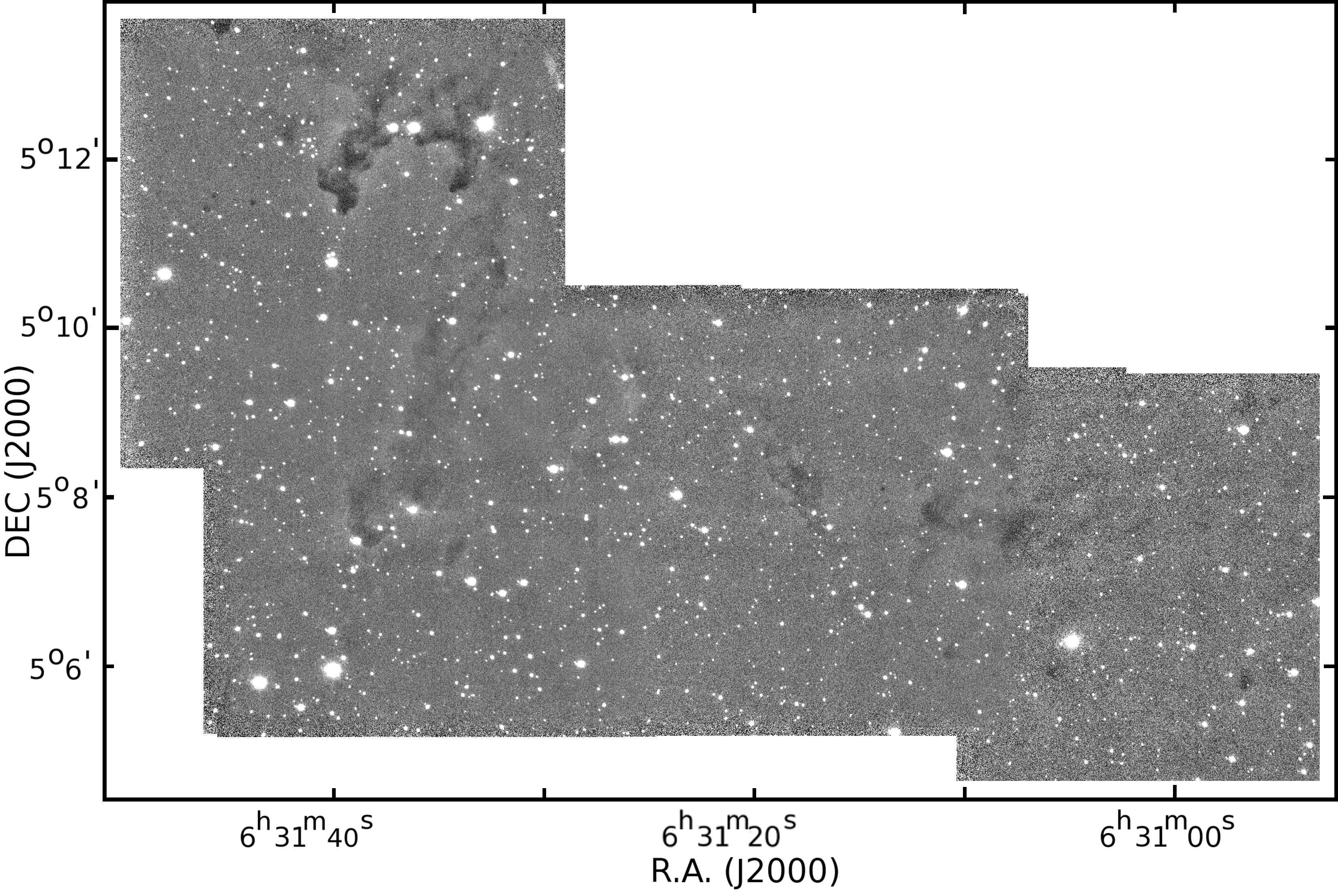}
    \caption{Fields 7, 8, 14, and 15 imaged in the NB 1.282 P$\beta$ filter.}
 \label{fig:pbmap}
 \end{figure}

\begin{figure}
 \centering
 \includegraphics [width=9cm]{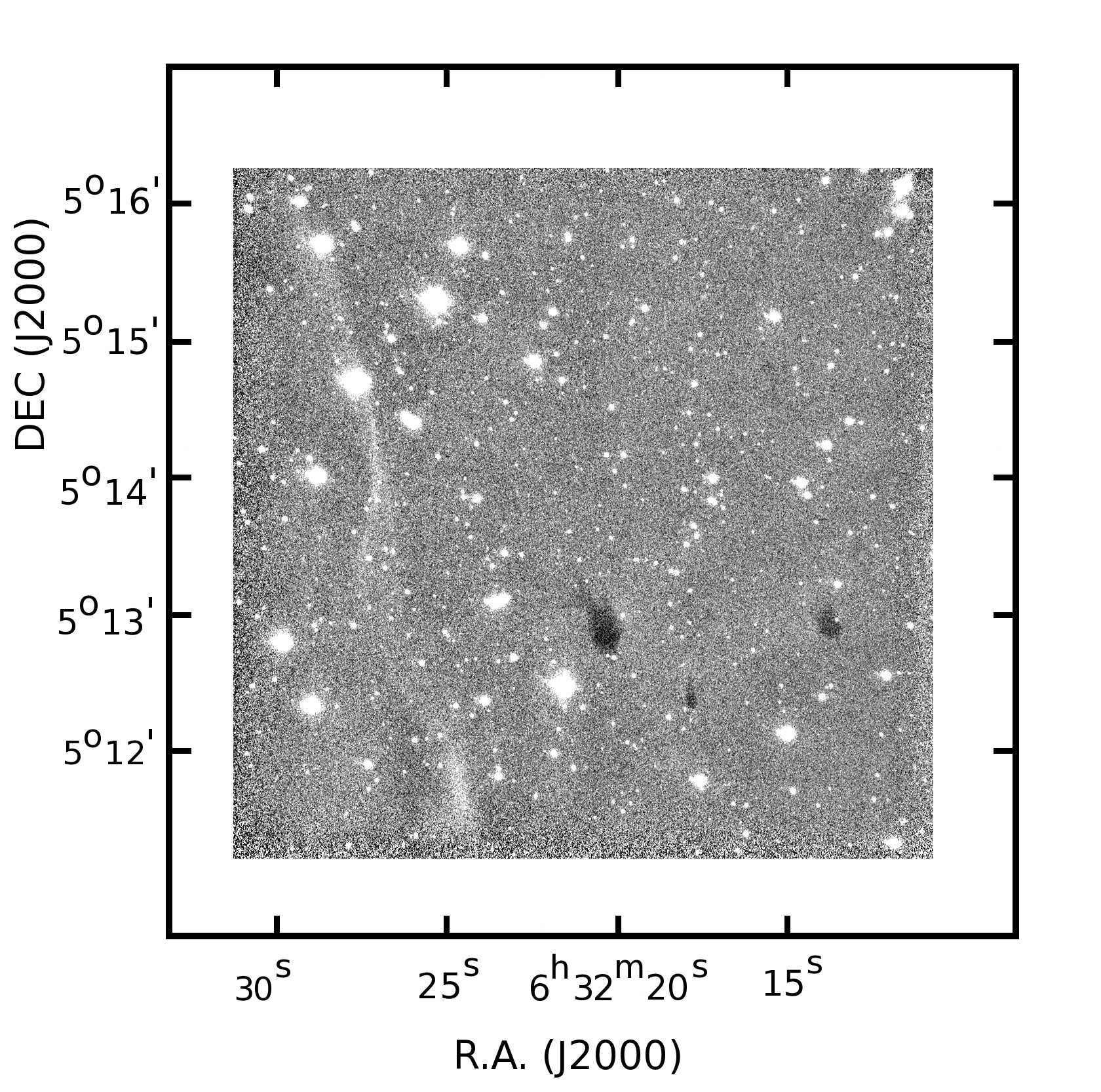}
    \caption{Field 19 imaged in the NB 1.282 P$\beta$ filter.}
 \label{fig:f19pbmap}
 \end{figure}

\begin{figure}
 \centering
 \includegraphics [width=9cm]{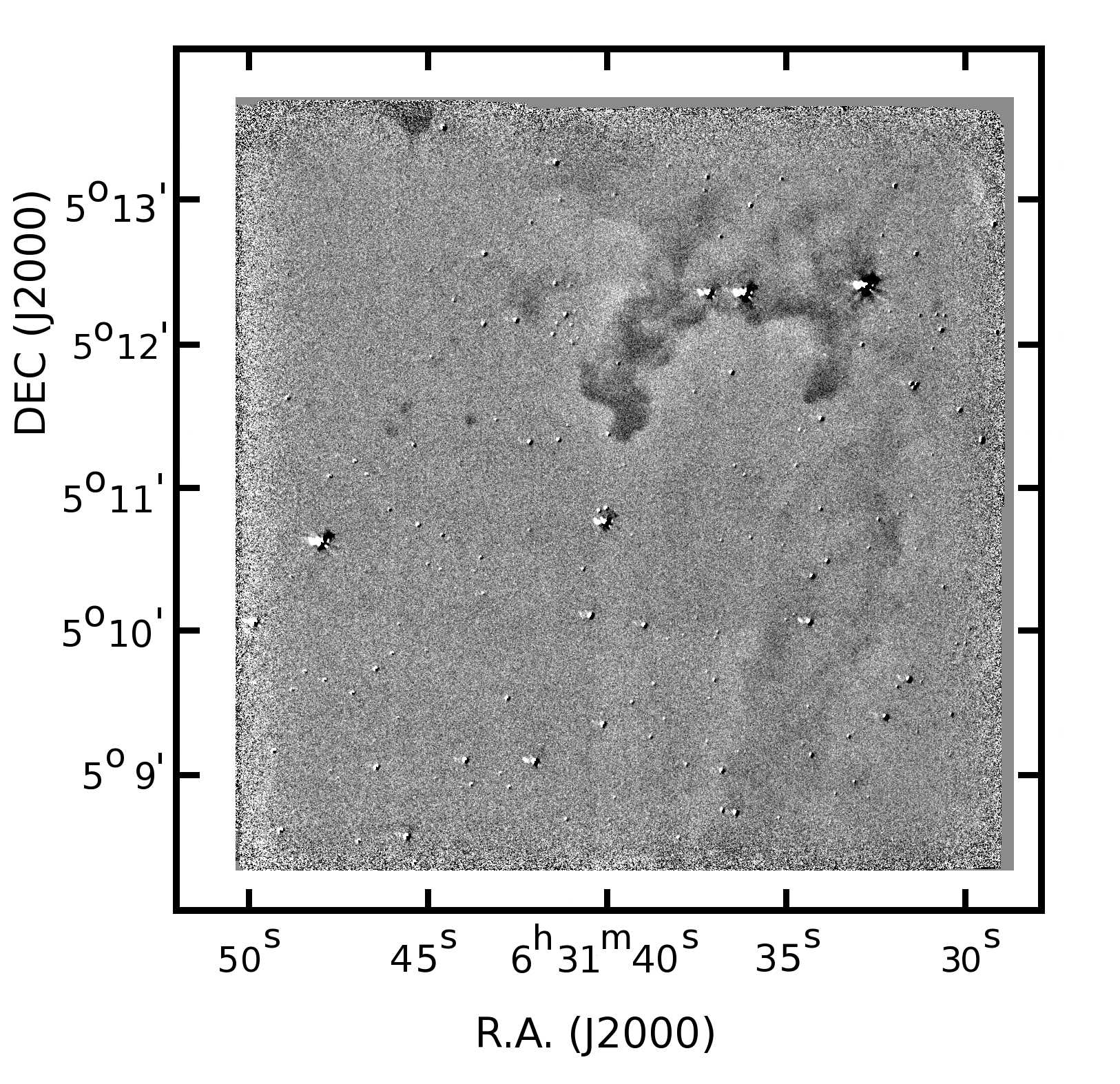}
    \caption{NB 1.282 P$\beta$ and NB 1.257 difference image (positive) of F15.}
 \label{fig:f15pbminuscontmap}
 \end{figure}

\subsection{NB 1.282 P$\beta$, NB 1.257, and NB 2.167 Br$\gamma$ imaging}
\label{sect:pbimages}

The combined NB 1.282 P$\beta$ image of fields 7, 8, 14, and 15, is shown in Fig. \ref{fig:pbmap} and of F19 in Fig. \ref{fig:f19pbmap}.
Some globulettes, shells, and both large elephant trunks are seen in absorption against the background. A NB 1.257 image of F15 is shown in Fig. \ref{fig:jcontmap}. This filter samples the continuum next to the P$\beta$ filter and no signs of the globulettes or trunks are seen. The continuum-subtracted P$\beta$ image is shown in Fig. \ref{fig:f15pbminuscontmap}, and we suggest that the P$\beta$ line emission causes the high background surface brightness seen in the \J\ and P$\beta$ images.

Column 4 of Table \ref{table:globulettes} indicates whether the globulette is detected in the NB 1.282 P$\beta$ image. Of the 84 inspected globulettes and elephant trunks, 14 appear clearly dark in the image. In the objects seen in absorption in P$\beta$, the absorption is strongest on the side that faces the central cluster. This is evident especially in the shells. The clearest example of absorption is seen in the Claw in F15.
Besides the Wrench and the Claw, also the globulettes in the String west of the Claw (RN 91, 93, 94 and 95) and especially the small globulette RN 88 are seen in absorption. The other globulettes that are seen as dark structures include RN 31, 35, 40, and the small RN 44, and globulettes RN 114, 122, and 129 in F19.

Positive P$\beta$ surface brightness is seen in a wispy filament west of the Wrench in Fig. \ref{fig:f15pbminuscontmap} at around 6:31:26, 5:09:00, and in a wider, N-S-directed filament east of RN 129 at R.A. of about 6:32:27 in F19.
As these positions coincide with bright regions in the H$\alpha$\ images in \citetalias{gahmetal07}, we interpret that these features are due to P$\beta$ emission in the background. Faint limb brightening is seen around the Claw and the western tip of the Wrench. The eastern tip of the Wrench has a faint but thin rim similar to the bright rims seen in \Htwo. Indication of surface brightness is seen also in the direction of the small-sized globulettes RN 78, 80, and 84 south of the Wrench.

The globulettes where absorption in P$\beta$ is detected, have radii larger than the globulette mean value of 2.5 kAU \citepalias{gahmetal07} and masses $>11$ M$_{Jup}$ which is higher than the typical globulette mass obtained in \citetalias{gahmetal07}. The globulettes showing P$\beta$ surface brightness are all smaller than the mean radius. The globulettes that are seen only through surface brightness comprise half of the Rosette Nebula globulettes where \citetalias{gahmetal07} detected bright rims in H$\alpha$ in the SOFI fields. 
Absorption in the NB 1.282 P$\beta$ image is also seen in the large shells, especially in F7 and F8, and stars in these areas appear very reddened. The absorption appears strongest just behind the \Htwo-bright rims.
An exception to this is in the dust shell in F7 where a maximum in the P$\beta$ absorption is seen farther away from the edge of the shell at 6:31:07.6, 5:07:36.
The Wrench has P$\beta$ absorption throughout it, even in the ``Handle'' of the Wrench which appears as more diffuse in the \jhks\ images.

The NB 2.167 Br$\gamma$ image of F14 is shown in Fig. \ref{fig:brgmap}. It has little signs of the elephant trunks or globulettes. Only faint surface brightness is seen along the \Htwo-bright rim in the tip of the western jaw of the Wrench and also around the star in the Cove.

\begin{table*}
 \caption{Globulette appearances in the NB 2.124 \Htwo\ S(1) and NB 1.282 P$\beta$ images. RN is the globulette number from \citetalias{gahmetal07}. The last two items refer to the eastern jaws of the Wrench and the Claw. \Htwo-rim indicates whether the globulette has a bright rim in the \Htwo\ 2.12 $\mum$ image. An H in this column indicates halo, a bright rim encircling the entire globulette. \Htwo-SB describes the globulette surface brightness at \Htwo\ 2.12 $\mum$ and P$\beta$ describes whether the globulette is seen in absorption at P$\beta$ (symbols: ({\textbf v}ery) {\textbf F}aint/{\textbf V}isible/{\textbf B}right/{\textbf D}ark/{SB} positive surface brightness). The last column gives number of stars visible in NIR behind the globulette. }
 \label{table:globulettes}
 \centering
 \begin{tabular}{c c c c c | c c c c c | c c c c c}
 \hline\hline
 RN & \Htwo-rim & \Htwo-SB & P$\beta$ & Star & RN & \Htwo-rim & \Htwo-SB & P$\beta$ & Star & RN & \Htwo-rim & \Htwo-SB & P$\beta$ & Star \\
 \hline
 31 &  x  &  V &  D & 1 &   66 &  -  &  V &  V & - &    115 &  x  &  B &  - & - \\
 32 &  -  &  - &  - & - &   67 &  -  &  B &  V & - &    116 &  x  &  B &  - & 2 \\
 33 &  -  & vF &  - & 1 &   68 &  x  &  V &  F & - &    117 &  x  &  V &  - & - \\
 34 &  -  &  F &  F & - &   69 &  -  &  - &  - & - &    118 &  -  &  F &  - & - \\
 35 &  x  &  V &  D & 2 &   70 &  -  &  B &  - & - &    119 &  x  &  B &  - & - \\
 36 &  x  &  V &  - & - &   72 &  -  & vF &  V & - &    120 &  x  &  B &  - & 1 \\
 37 &  x  &  V & vF & 1 &   73 &  x  &  B &  F & 1 &    121 &  x  &  V &  - & - \\
 38 &  x  &  B & vF & 1 &   74 &  -  &  V &  - & - &    122 &  x  &  B &  D & 1 \\
 39 &  x  &  V &  - & 1 &   75 &  x  & vB &  - & - &    123 &  x  &  V &  - & - \\
 40 &  x  &  V &  F & - &   76 &  -  &  F &  - & - &    124 &  x  &  V &  - & - \\
 41 &  -  &  F &  - & - &   77 &  x  &  F &  F & - &    125 &  x  &  V &  - & - \\
 42 &  x  & vF &  - & - &   78 &  H  &  B & SB & - &    126 &  x  &  B &  - & - \\
 44 &  x  &  F &  D & - &   79 &  x  &  B &  - & 1 &    127 &  x  &  F &  - & - \\
 46 &  x  &  F &  - & - &   80 &  x  &  B & SB & - &    128 &  -  &  - &  - & - \\
 47 &  x  &  B &  - & 1 &   84 &  x  &  B & SB & - &    129 &  x  &  B &  D & 3 \\
 48 &  x  &  V &  - & 1 &   85 &  -  &  - &  - & - &    130 &  -  &  F &  - & - \\
 49 &  x  &  B &  F & - &   86 &  x  &  F &  - & - &    131 &  -  &  - &  - & - \\
 50 &  -  &  V &  - & 1 &   87 &  x  &  B &  - & - &    132 &  x  & vF &  - & - \\
 51 &  -  &  B &  - & - &   88 &  x  &  B &  D & - &    133 &  x  &  B &  - & - \\
 52 &  x  &  F &  - & 1 &   89 &  x  &  F &  - & - &    134 &  -  &  V &  - & - \\
 53 &  -  &  F &  F & - &   90 &  x  &  F &  - & - &    135 &  -  &  F &  - & - \\
 54 &  -  & vF &  - & 1 &   91 &  x  &  B &  F & 1 &    136 &  x  &  F &  - & - \\
 55 &  x  &  B &  - & 1 &   92 &  x  &  V &  - & 1 &    137 &  x  &  F &  - & - \\
 56 &  -  &  B &  - & - &   93 &  x  &  B &  F & - &    138 &  -  &  V &  - & - \\
 57 &  -  &  F &  - & 2 &   94 &  H  &  B &  D & 2 &    139 &  x  &  F &  - & - \\
 58 &  -  &  V &  - & - &   95 &  x  &  B &  D & 1 &   Triangle\tablefootmark{a} & x & - & - & - \\
 59 &  x  &  B &  - & - &   113 &  x  &  B &  - & 1 &   Wrench-E & x & B & D & x \\
 65 &  -  &  F &  - & - &   114 &  x  &  B &  D & 1 &   Claw-E   & x & V &  D & x \\
 \hline
 \end{tabular}
 \tablefoot{
 \tablefoottext{a}{ Not listed in \citetalias{gahmetal07}. Center coordinates 6:31:40.63, 5:06:44.95 (J2000.0).}
 }
 \end{table*}



 \subsection{Stellar \jhks\ photometry}
 \label{sect:photometry}

\begin{figure}
 \centering
\includegraphics[width=7cm]{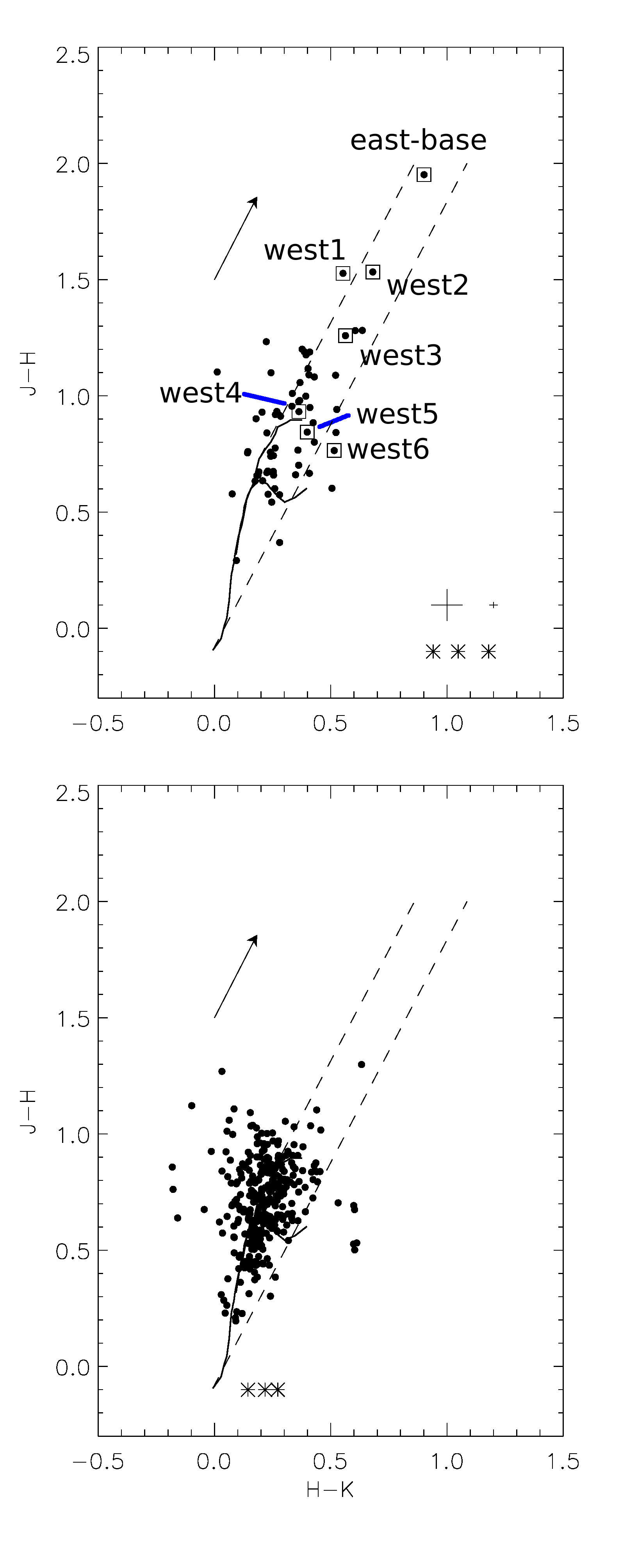}
 \caption{The \J$-$\H, \H$-$\Ks\ color-color diagram of the shell (upper panel) and background objects (lower panel) in F14. Filled circles mark the observed stars and boxes mark the stars behind the globulettes. Asterisks mark stars with only \H$-$\Ks\ colors. The crosses in the bottom right-hand corner indicate the maximum (left) and the average error (right). A reddening vector of 3\umag\ is indicated. The names refer to the eastern and western jaw of the Wrench. The numbering starts from the tip of the jaw. The corresponding coordinates are (in J2000.0) (1): 6:31:34.49, 5:07:01.8, (2): 6:31:34.44, 5:07:07.6, (3): 6:31:33.60, 5:07:32.0, (4): 6:31:34.03, 5:07:41.4, (5): 6:31:33.34, 5:07:40.4, (6): 6:31:32.95, 5:07:45.2. East-base refers to the most reddened star in the base of the eastern jaw. Its coordinates are 6:31:38.06, 5:08:06.0 (J2000.0).
}
 \label{fig:jhhkdiagramf14}
 \end{figure}

 In order to study the large-scale extinction (see Sect. \ref{sect:av}), each imaged SOFI field was divided into two areas, one covering major shell features and one covering surrounding regions (background areas). The globulettes were treated individually. The \J$-$\H, \H$-$\Ks\ color-color diagrams of the shell and background in F14 are shown in Fig. \ref{fig:jhhkdiagramf14}. Similar figures of other observed fields are in Figs. \ref{fig:jhhkdiagramf7}-\ref{fig:jhhkdiagramf19}. The upper panel shows stars located in the shell region, and the lower panel those located in the field. Filled circles mark the observed stars, and boxes mark stars in the direction of globulettes. Asterisk is used for stars detected only in the \H\ and \Ks\ bands (see Sect. \ref{sect:globav} for details).
The thick lines mark the main and giant sequences defined by \citet{bessellbrett88} (hereafter \citetalias{bessellbrett88}) and the dashed lines show the reddening lines according to \citetalias{bessellbrett88}. Reddened stars without IR excess lie between the reddening lines.

In the shell region most of the observed stars are reddened. The background region displays little reddening and the stars mostly follow the main sequence. High reddening is observed in stars which most probably are behind the globulettes. Some of the highest $A_{V}$ stars in F14 are marked in Fig. \ref{fig:jhhkdiagramf14}. The highest $A_{V}$ is detected in the base of the eastern jaw of the Wrench where other reddened stars are also seen. The high-$A_{V}$ stars in the western jaw are marked in the diagram and studied further in Sect. \ref{sect:globav}. The highest reddening detected in the shell \J$-$\H, \H$-$\Ks\ diagrams corresponds to almost 17\umag\ and it is located in the direction of the densest P$\beta$ core in F7.

Even though the majority of the objects plotted in Fig. \ref{fig:jhhkdiagramf14} and Figs. \ref{fig:jhhkdiagramf7}-\ref{fig:jhhkdiagramf19} lie close to the unreddened main sequence or near and between the reddening lines, a large number of objects are seen farther out to the left or right of the reddening lines. The selection rules applied to the stellar catalog (Sect. \ref {sect:datareduction}) already removed most of such deviating objects but some still remain. The nature of such outliers has been discussed in detail in \citet{fosteretal08}. They conclude that besides genuine NIR excess stars the outliers can be produced by evolved stars, brown dwarfs, binaries, quasars, AGNs and redshifted (z=0.1--0.5) galaxies. The Rosette Nebula is located in the Galactic anti-center where the interstellar extinction in the field is moderate and therefore contamination by extragalactic sources is not unlikely. The bright nearby galaxies can be spatially resolved as extended sources but not necessarily the distant faint ones.

Most of the objects located below the lower reddening line have m$_{Ks}>18$\umag\ and they are not seen at all in the {\textit Spitzer} IRAC images or are detected only as faint objects in the IRAC 3.6 and/or 4.5 $\mum$ band. Only one of the objects is seen in the 8.0 $\mum$ image. Because they are faint in IRAC, we cannot compute their IRAC colors and confirm whether they have IR excess or suffer from NIR measurement errors. Regardless of being real objects or extreme measurement errors the reddenings derived from these outliers are meaningless and will not be considered in the following.

 \subsection{Globulette \jhks\ and $\Htwo$\ surface brightness}

We selected globulettes and trunks with distinct bright rims and some with large, denser cores to study their \Htwo\ 2.12 $\mum$ surface brightness.The desired surface brightness values were derived from each of the \J, \H, \Ks, and \Htwo\ filter images in the same manner. We computed the surface brightness in detector counts from the mean pixel value of the core/rim, and subtracted a mean pixel value of an empty nearby background. This value is converted into (2MASS) magnitudes per square arcsecond. This is used to obtain the surface brightness in S10 units (1 S10 is defined as 27.78 mag/square arcsec) which  is transferred to W m$^{-2}$ sr$^{-1} \mum^{-1}$ using Table 2 of \citet{leinertetal98}.

To get the globulette rim and core magnitudes in the \J, \H, and \Ks\ filters to the 2MASS system, a scaling factor was derived for each SOFI \jhks\ field and filter using the instrumental stellar fluxes and their 2MASS magnitudes. The fields were then scaled before extracting the pixel values from the rims and cores, and no further zero-point correction or 2MASS transformation formulae are thus needed. The absolute values of the scaling factors range from 0.74 to 1.44 in \jhks. Because they were derived using the 2MASS magnitudes, the errors and rms are very small (<$10^{-2})$. 

For the NB 2.124 \Htwo\ S(1) images the conversion into the 2MASS system was done as follows. The images were scaled to the \Ks\ images, which were already in the 2MASS system, using common unreddened field stars. The scaling factor values are $\sim$4.6 which includes a factor of three from the \Htwo/\Ks\ integration time ratio. The errors in different fields range from 0.02 to 0.04 with a typical rms of 0.1. After this the \Ks\ and NB 2.124 \Htwo\ S(1) images give the same counts for stars, which is not physically possible. 
The NB 2.124 \Htwo\ S(1) images were then scaled by multiplying them by the ratio T(NB 2.124)/T(\Ks) where T is the transmission integrated over the filter. The wavelength-dependent SOFI filter transmission curves are available at ESO's website\footnote{\texttt{http://www.eso.org/sci/facilities/lasilla/\allowbreak{}instruments/sofi/inst/Imaging.html}}.

Tables \ref{table:coresb} and \ref{table:rimsb} list the obtained \jhks\ and \Htwo\ 2.12 $\mum$ surface brightnesses in the cores and rims of several globulettes, respectively. RN 39 has a dark artifact in the core in the \Ks\ image, so the measured surface brightness value is not reliable. The mean \Htwo/\Ks\ surface brightness ratio for the globulette cores in Table \ref{table:coresb} is 0.26$\pm0.02$ with an rms of 0.09. For the rims in Table \ref{table:rimsb} these values are 0.30$\pm0.01$ and 0.07. No positive surface brightness is detected in the \Js\ band images in the cores of those globulettes that are seen in absorption in the NB 1.282 P$\beta$ image.

The narrow-band NB 2.124 \Htwo\ S(1) images were obtained in jittering mode which smears the background toward zero level. Therefore the surface brightness computed from the jittered images is a differential value and gives a lower limit for the \Htwo\ 2.12 $\mum$ surface brightness. The measured core surface brightness in the \Htwo\ 2.12 $\mum$ line becomes unreliable in the largest globulette RN 129 where the globulette size is approximated with an ellipse with semi-axes of 26\arcsec\ and 9.4\arcsec \citepalias{gahmetal07}. The semimajor axis approaches the jitterbox width (30\arcsec), and the core of the globulette goes darker due to the contrast effect of the data reduction process. However, the computed rim brightnesses of the on-off and jittering images do not differ significantly.

The result shows that the observed \Htwo\ 2.12 $\mum$ line can only account for 20-45\% of the \Ks\ band surface brightness emission in the globulette rims (see column 5 of Tables \ref{table:coresb} and \ref{table:rimsb}). This confirms the results of G13 and the broader survey here does not find any correlation between the globulette location or size and the observed \Htwo/\Ks\ surface brightness ratio. The P$\beta$-dark cores also have no systematic similarities.
Because the continuum and Br$\gamma$ images suggest that the observed emission from the globulettes is line emission, other \Htwo\ lines inside the \Ks\ band should contribute to the observed \Ks\ flux. Also, the bright rims detected in the \J\ and \H\ bands can be explained via these other \Htwo\ lines.

 \begin{table*}
 \caption{Surface brightness of selected globulette cores. Column 1 lists the globulette ID. Columns 2-5 contain the surface brightness measured in the \J, \H, \Ks, and \Htwo\ 2.12 $\mum$ bands, respectively. The surface brightness unit is W m$^{-2}$ sr$^{-1} \mu$m$^{-1}$. 
 The last column is the ratio between the  \Htwo\ and \Ks\ surface brightnesses.} 
 \label{table:coresb}
 \centering
 \begin{tabular}{c c c c c c | c c c c c c}
 \hline\hline
 RN & I(\J) & I(\H) & I(\Ks)  & I(\Htwo) & f &  RN & I(\J) & I(\H) & I(\Ks)  & I(\Htwo) & f \\
 \hline

   31 &   \ldots &  3.1E-08 &  3.1E-08 &  9.5E-09 & 0.30 &   73 &  2.2E-08 &  1.1E-07 &  8.7E-08 & 1.7E-08 & 0.20 \\
   34 &  1.9E-08 &  5.7E-08 &  6.7E-08 &  1.8E-08 & 0.26 &   77 &  1.1E-07 &  6.1E-08 &  4.7E-08 &  1.0E-08 & 0.22 \\
   35 &   \ldots &  6.7E-08 &  3.6E-08 &  1.0E-08 & 0.28 &   78 &  2.2E-07 &  1.6E-07 &  1.4E-07 &  3.0E-08 & 0.21 \\
   38 &   \ldots &  5.2E-08 &  2.7E-08 &  1.3E-08 & 0.48 &   79 &  2.3E-07 &  2.0E-07 &  1.7E-07 &  2.5E-08 & 0.14 \\
   39 &  8.6E-08 &  9.9E-08 &  7.5E-08: &  1.5E-08 & 0.19: &   84 &  2.1E-07 &  1.9E-07 &  1.6E-07 &  3.2E-08 & 0.20 \\
   40 &  2.8E-08 &  7.8E-08 &  3.1E-08 &  7.4E-09 & 0.23 &   88 &   \ldots &  9.9E-08 &  9.1E-08 &  1.7E-08 & 0.19 \\
   44 &   \ldots &  7.3E-09 &  4.0E-08 &  1.5E-08 & 0.37 &   95 &   \ldots &  3.4E-08 &  4.0E-08 &  2.1E-08 & 0.52 \\
   47 &  1.5E-07 &  1.7E-07 &  1.4E-07 &  3.40-08 & 0.21 &  114 &   \ldots &  1.1E-07 &  6.5E-08 &  1.4E-08 & 0.21 \\
   49 &  6.8E-08 &  6.9E-08 &  7.4E-08 &  2.3E-08 & 0.31 &  122 &   \ldots &  1.2E-08 &  3.1E-08 &  7.5E-09 & 0.24 \\
   50 &  1.8E-07 &  1.1E-07 &  1.1E-07 &  2.5E-08 & 0.23 &  129 &   \ldots &  1.4E-07 &  2.5E-08 &  4.4E-09 & 0.18 \\
   55 &  8.6E-08 &  1.3E-07 &  1.0E-07 &  2.8E-08 & 0.28 & Claw-E &  \ldots &  4.4E-08 &  7.0E-08 &  1.8E-08 & 0.26 \\

 \hline
 \end{tabular}
 \end{table*}

 \begin{table*}
 \caption{ As Table 3 for surface brightness of selected globulette rims.}
 \label{table:rimsb}
 \centering
 \begin{tabular}{c c c c c c | c c c c c c}
 \hline\hline
 RN & I(\J) & I(\H) & I(\Ks) & I(\Htwo) & f &  RN & I(\J) & I(\H) & I(\Ks) & I(\Htwo) & f \\
 \hline

      31 &  1.1E-07 &  9.1E-08 &  1.1E-07 &  2.5E-08 & 0.23 &    79 &  4.2E-07 &  2.9E-07 &  3.0E-07 &  6.8E-08 & 0.23 \\
      35 &  1.5E-07 &  1.1E-07 &  1.4E-07 &  3.4E-08 & 0.24 &    80 &  3.5E-07 &  2.4E-07 &  2.5E-07 &  5.7E-08 & 0.23 \\
      38 &  1.1E-07 &  9.1E-08 &  1.2E-07 & 3.9E-08 & 0.33 &    84 &  4.3E-07 &  2.4E-07 &  2.8E-07 &  6.7E-08 & 0.23 \\
      39 &  1.1E-07 &  9.8E-08 &  1.4E-07 &  3.8E-08 & 0.27 &    86 &  9.6E-08 &  6.6E-08 &  1.1E-07 &  3.5E-08 & 0.32 \\
      40 &  2.0E-07 &  1.5E-07 &  1.6E-07 &  6.0E-08 & 0.39 &    88 &  3.2E-08 &  7.0E-08 &  1.3E-07 &  4.1E-08 & 0.32 \\
      41 &  8.4E-08 &  4.1E-08 &  8.2E-08 &  1.8E-08 & 0.22 &    90 &  2.8E-08 &  5.5E-08 &  7.2E-08 &  2.9E-08 & 0.40 \\
      42 &   \ldots &  4.0E-08 &  7.7E-08 &  2.5E-08 & 0.33 &    91 &   \ldots &  7.6E-08 &  9.3E-08 &  4.1E-08 & 0.44 \\
      44 &  3.5E-08 &  1.1E-07 &  9.3E-08 &  3.7E-08 & 0.39 &    93 &  7.4E-08 &  5.7E-08 &  1.4E-07 &  5.7E-08 & 0.40 \\
      47 &  2.5E-07 &  2.1E-07 &  2.4E-07 &  7.6E-08 & 0.31 &    94 &  1.2E-07 &  1.2E-07 &  1.6E-07 &  4.6E-08 & 0.29 \\
      49 &  9.0E-08 &  1.0E-07 &  1.6E-07 &  3.4E-08 & 0.21 &    95 &  2.0E-07 &  1.7E-07 &  2.5E-07 &  6.9E-08 & 0.27 \\
      51 &  4.7E-08 &  2.6E-08 &  7.9E-08 &  2.7E-08 & 0.34 &   114 &  1.7E-07 &  2.3E-07 &  2.3E-07 &  6.7E-08 & 0.29 \\
      52 &  2.0E-07 &  1.0E-07 &  1.8E-07 &  5.2E-08 & 0.29 &   122 &  6.3E-08 &  9.4E-08 &  1.1E-07 &  5.0E-08 & 0.44 \\
      55 &  2.0E-07 &  1.6E-07 &  2.2E-07 &  7.5E-08 & 0.34 &   125 &  2.2E-08 &  1.0E-07 &  1.3E-07 &  4.0E-08 & 0.30 \\
      73 &  2.8E-07 &  1.5E-07 &  1.8E-07 &  5.3E-08 & 0.29 &   129 &  2.5E-07 &  2.6E-07 &  2.3E-07 &  5.0E-08 & 0.21 \\
      75 &  1.8E-07 &  1.1E-07 &  1.8E-07 &  5.6E-08 & 0.30 & Wrench-E &  4.6E-07 &  3.5E-07 &  3.3E-07 &  6.5E-08 & 0.20 \\
      77 &  1.8E-07 &  9.2E-08 &  1.8E-07 &  5.5E-08 & 0.31 & Wrench-Cove &  5.1E-07 &  3.4E-07 &  3.9E-07 &  9.6E-08 & 0.25 \\
      78 &  3.0E-07 &  1.8E-07 &  2.1E-07 &  6.7E-08 & 0.31 &  Claw-E &  2.2E-07 &  1.7E-07 &  2.3E-07 &  6.2E-08 & 0.22 \\

\hline
 \end{tabular}
 \end{table*}


%

\section{Discussion} \label{sect:discussion}

In our previous study of the molecular line emission from globulettes \citepalias{gahmetal13}, it was concluded that the most massive objects contain dense cores, but that the density is high also close to the surface of the objects and the gas is molecular all the the way to the surface. As demonstrated in Sect. \ref{sect:pbimages}, the elephant trunks, shells, and some of the globulettes appear as dark against the nebular background in P$\beta$ and \J\ band images. In the shells the density is high right behind the bright rims and falls with distance from them.

The dense cores are not evident in the H$\alpha$ images discussed in \citetalias{gahmetal07}, and below we will explore whether these objects contain more mass than estimated in the optical survey. It should be emphasized that most of the globulettes discussed in this section are more massive than the majority of the Rosette globulettes in \citepalias{gahmetal07} where the masses are typically $<13$ M$_{Jup}$.
The less massive objects are transparent in NIR and are detected only through faint surface brightness in the NB 2.124 \Htwo\ S(1) and \Ks\ images. Moreover, the spatial resolution of SOFI is not high enough to enable detailed study of the smallest globulettes.

We will use visual extinction, $A_{V}$, as a tool to study the large-scale structure in the SOFI fields and in individual stars behind globulettes. This will yield information on the density along the line-of-sight. We have also extracted {\textit Spitzer} MIR data to identify stars in formation and to better constrain the excitation mechanisms for the \Htwo\ 2.12 $\mum$ emission, and Herschel FIR data to independently study the globulette densities.

\subsection{Visual extinction}
 \label{sect:av}

We use the NICER method described in \citet{nicer_lombardietalves} to produce an $A_{V}$ map of the imaged region. NICER is a convenient method to study the large-scale extinction structure, but it can not be used to study the small-scale structure in positions such as the globulettes where only one or two stars are detected. The approach is statistical and the reliability and spatial resolution depend on the stellar density.

The final combined \jhks\ photometry catalog contains objects that have been detected in all three bands. The NICER method can however also utilize objects with only \H$-$\Ks\ colors to estimate their $A_{V}$. For this, we retrieved such objects from the original non-filtered catalogs and used the selection rules to filter out non-stellar objects. Altogether 13 stars remain and their observed SOFI \H$-$\Ks\ colors were transformed into 2MASS using the conversion formulas in \citet{suutarinenetal13} before adding them to the NICER catalog.
The catalogs that contain the SOFI \jhks\ photometry for the ON and the OFF regions are available as Tables 4 and 5 at the CDS.

\addtocounter{table}{2}

\begin{figure}
 \resizebox{\hsize}{!}{\includegraphics{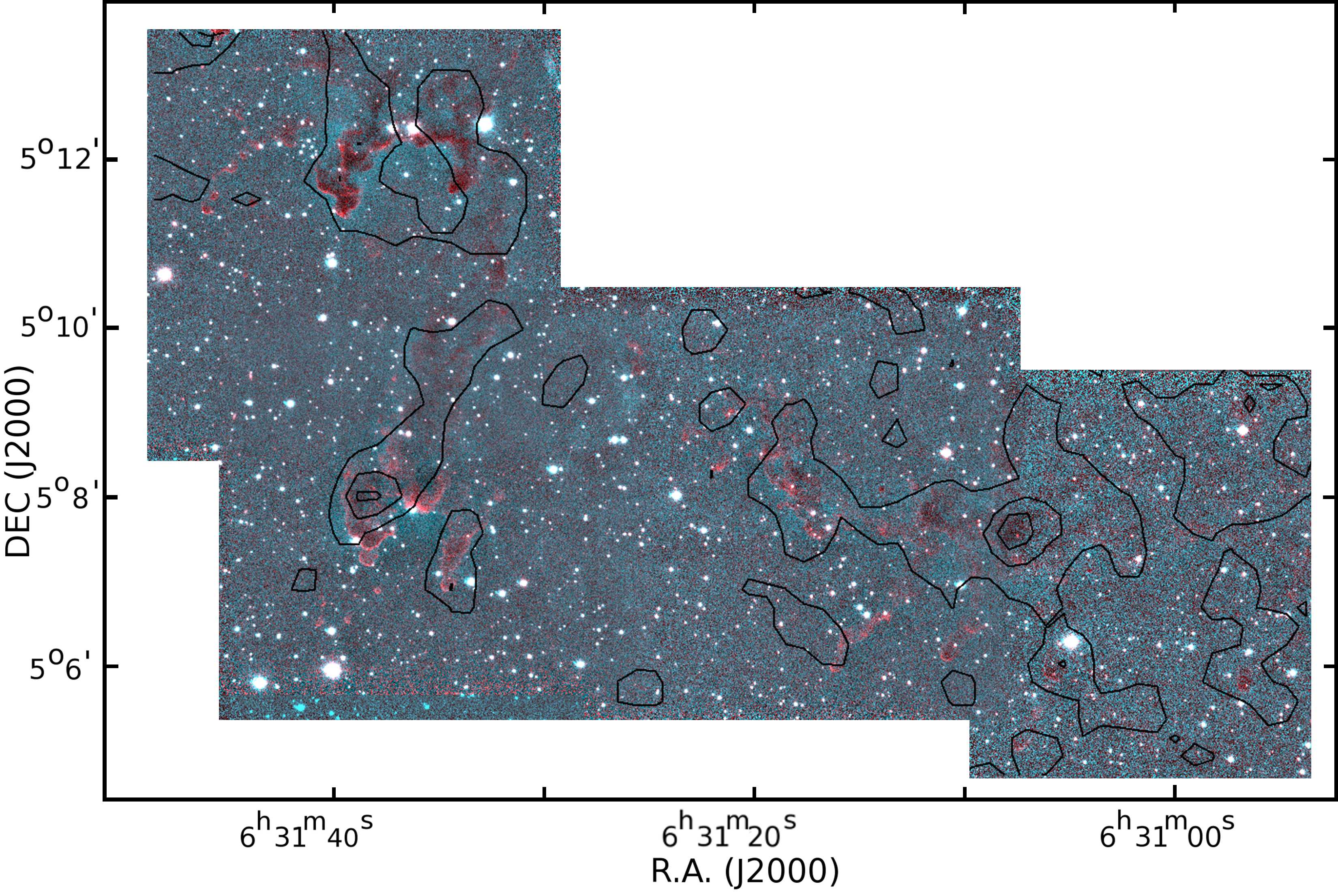}}
    \caption{Visual extinction in the Rosette Nebula. $A_{V}$ contours of 2, 4, and 6 magnitudes are shown on a false-color image. NB 2.124 \Htwo\ S(1) image is coded in red and NB 1.282 P$\beta$ image in turquoise.}
 \label{fig:avcontours}
 \end{figure}

The Gaussian used for smoothing the data has a FWHM of 26.4\arcsec\ and the pixel size of the NICER $A_{V}$ map is 13.2\arcsec. The resolution is limited by the stellar density in the photometry catalog as using a higher resolution will make gaps appear in the map. The empty pixels with no more than two empty neighbors were interpolated using the average value of the neighboring pixels that remain. The \citetalias{bessellbrett88} reddening law was assumed. 
The $A_{V}$ contours are superposed on the false-color NB 2.124 \Htwo\ S(1)-NB 1.282 P$\beta$ image of the four overlapping fields in Fig. \ref{fig:avcontours}. The $A_{V}$ peaks correspond to the areas where the P$\beta$ absorption is the strongest, e.g. the Wrench and the Claw and the dust shell between F7 and F8. Most of the globulettes are not seen as the resolution of the $A_{V}$ map is too low as no stars are detected in their direction. Two globulettes in F7, RN 31 and RN 35, are traced by the $A_{V}$ contours to some degree.

Outside the shells, the typical $A_{V}$ in the map ranges from $\sim$1\umag\ to 1.5\umag. In the shell region, the extinction is typically twice this. The highest $A_{V}$ peaks are $6.3$\umag\ in the Wrench and $8.6$\umag\ in the dust shell in F7.

Most of the globulettes have sizes on the order of 1\arcsec\ to 10\arcsec\ and about a third of the globulettes screens stars but even then there at the most 2-3 stars behind them (see column 5 of Table \ref{table:globulettes}). Larger globulettes have a higher probability to screen background stars, which will skew the resulting mass estimates from this type of study toward higher than typical globulette masses.
The $A_{V}$ in the globulettes can be estimated in the traditional manner using the observed NIR colors of individual stars seen in their direction. This method for estimating the extinction in the globulettes does not work for $A_{V}$ higher than $\sim$15\umag\ because the late main sequence stars at the distance of the Rosette Nebula will fall below the detection limits. Stars first disappear in \J, then in \H, and finally in \Ks\ when $A_{V}$ increases.

\subsubsection{Individual globulettes}
\label{sect:globav}

The NOT H$\alpha$ images have been used to identify stars behind the globulettes. If a star is clearly seen in the H$\alpha$ image it is classified as a foreground star. Stars faint or not detected in H$\alpha$ have been considered as background stars. The stars well outside the reddening lines of the \J$-$\H, \H$-$\Ks\ diagram have been dismissed. The locations of remaining stars have been noted in the \J$-$\H, \H$-$\Ks\ diagram and an estimate for the $A_{V}$ has been extracted by measuring the distance between the star and the dwarf star main sequence (we assume that the background stars are not giants due to low observed magnitudes and Rosette's location towards the direction of the Galactic anti-center).
In addition, the observed \jhks\ magnitudes have been used to differentiate between the redder late (M type) main sequence branch and the earlier spectral type branch if possible. If a determination between the main sequence branches could not be made, a mean of the two $A_{V}$ values was used. Reference stars near the globulettes were used to estimate the $A_{V}$ contribution from the foreground.

Table \ref{table:globav} contains the estimated $A_{V}$ for the stars behind the globulettes/trunks. Column 1 lists the globulette ID and column 2 the $A_{V}$.
In RN 35, two stars were observed, one close to the head and one in the more diffuse tail. The last six items in Table \ref{table:globav} refer to the elephant trunks.
The naming of the objects in the Wrench (F14) is according to Fig. \ref{fig:jhhkdiagramf14}: F14-west describes the western jaw of the Wrench from the tip to the base and F14-east is located in the base of the eastern jaw of the Wrench. F15-tip refers to a star in the partially imaged trunk tip $\sim 2.5$\arcmin\ NE of the Claw. F15-tip has the highest  estimated $A_{V}$.
The trunk is seen at the northern edge of Fig. \ref{fig:all4jhk}. Column 3 of Table \ref{table:globav} lists the estimated \Htwo\ number density, $n$(\Htwo). We assumed that the linear line-of-sight size for globulettes is the average of the minor and major axes listed in \citetalias{gahmetal07}. For trunks, it is equal to the apparent width of the trunk at the position of the star.
The \citet{bohlinetal78} relation between the column density of molecular hydrogen, $N$(\Htwo), and optical extinction, $N$(\Htwo)$=0.94\times10^{21}$cm$^{-2}$mag$^{-1}$, is also 
used. Column 4 contains the computed globulette mass assuming a constant number density and a mean molecular mass of 2.8 amu per \Htwo\ molecule. The so-estimated masses are similiar to the optically derived masses in \citetalias{gahmetal07} only for the two smallest globulettes RN 33 and 54. The estimates for the larger globulettes are factors 7 to 15 times larger than the optically derived masses, which may be related to the fact that the larger objects contain small and very dense cores as shown in Sect. \ref{sect:photometry}. It should be pointed out that the numbers derived from the NIR data are based on extinction along only one line-of-sight through the objects and assuming constant density and the sizes listed in \citetalias{gahmetal07}. The typical number density in the globulettes and trunks is $\sim$a few$\times10^{4}$ cm$^{-3}$.

\begin{table}
 \caption{Properties of the globulettes as derived from background stars. The first column lists the globulette ID. The second column is the $A_{V}$ caused by globulettes and trunks. Columns 3 and 4 are the calculated number density and the mass. The last six entries refer to stars behind the densest trunks. Column 5 contains the optically derived masses by \citetalias{gahmetal07}.}
 \label{table:globav}
 \centering
 \begin{tabular}{r r c c c}
 \hline\hline
 RN & $A_{V}$ & $n$(\Htwo)  & M$_{NIR}$ & M$_{OPT}$   \\
    & (mag) & ($10^{4}$ cm$^{-3}$) & (M$_{Jup}$) & (M$_{Jup}$) \\
 \hline
   33 & 0.9 & 0.8 & 18.6 & 12.7 \\
   35 & 8.4 & 2.8 & 1164 & 172\\
      & 1.9 & -   & - & -\\
   52 & 8.6 & 4.5 & 504 & 34.7 \\
   54 & 1.6 & 1.8 & 18.6 & 5.1\\
   55 & 4.5 & 1.5 & 514 & 67.2 \\
   F14-west2 & 7.5 & 2.6 & - & - \\
   F14-west3 & 6.4 & 2.2 & - & - \\
   F14-west4 & 2.4 & 0.8 & - & - \\
   F14-west5 & 1.2 & 0.4 & - & - \\
   F14-east & 10.2 & 2: & - & - \\
   F15-tip & 14.2 & 4: & - & - \\
 \hline
 \end{tabular}
 \end{table}

RN 35 can be used to roughly evaluate the size and mass of the dense NIR core in a globulette. It has a small core seen in absorption in the \J\ band and two background stars
detected in NIR. They are 4.5\arcsec\ and 10.5\arcsec\ from the rim, and have visual extinctions 8.4\umag\ and 1.9\umag, respectively.
In the optical study \citetalias{gahmetal07} obtained semi-axes of 7.4\arcsec\ and 5.9\arcsec\, and a mass of 172 M$_{Jup}$. If a new ellipse is fit by hand to the NIR core that excludes the more diffuse H$\alpha$ tail, we obtain axes of 6.2\arcsec\ and 3.5\arcsec. This ellipse gives a NIR mass estimate of 578 M$_{Jup}$ for the core, about half the value computed in Table \ref{table:globav}. The \J\ band core that is seen in absorption gives an estimate of 256 M$_{Jup}$ which is about 1.5 times the optical mass. Even after correcting for the size of the densest core, the NIR mass estimate suggests higher masses than the optical estimate.

We can probe the $A_{V}$ distribution in the dense tip of eastern jaw of the Claw using two heavily reddened stars that are not included in the photometry catalog.
A star at the core of the tip is not detected in \J\ and is barely visible in the \H\ band, but it has a good signal in \Ks. Using the SOFI \Ks\ magnitude, we can estimate the lower limit of the A$_{V}$. We assume that the star is located between the reddening lines and that its spectral class is roughly M1. Using the \citetalias{bessellbrett88} reddening law gives $A_{V}\sim$10\umag\ at the distance of the Rosette Nebula. Earlier spectral type than M1 would increase the $A_{V}$. 
Another reddened star is detected toward the outer edge, 15\arcsec\ NE from the star inside the core. This star is detected at \H\ and \Ks\ and assuming spectral class M1, the $A_{V}$ is 12\umag.

\subsection{Spitzer imaging: dust composition}
 \label{sect:spitzer}

\begin{figure}
 \centering
 \includegraphics [width=8cm]{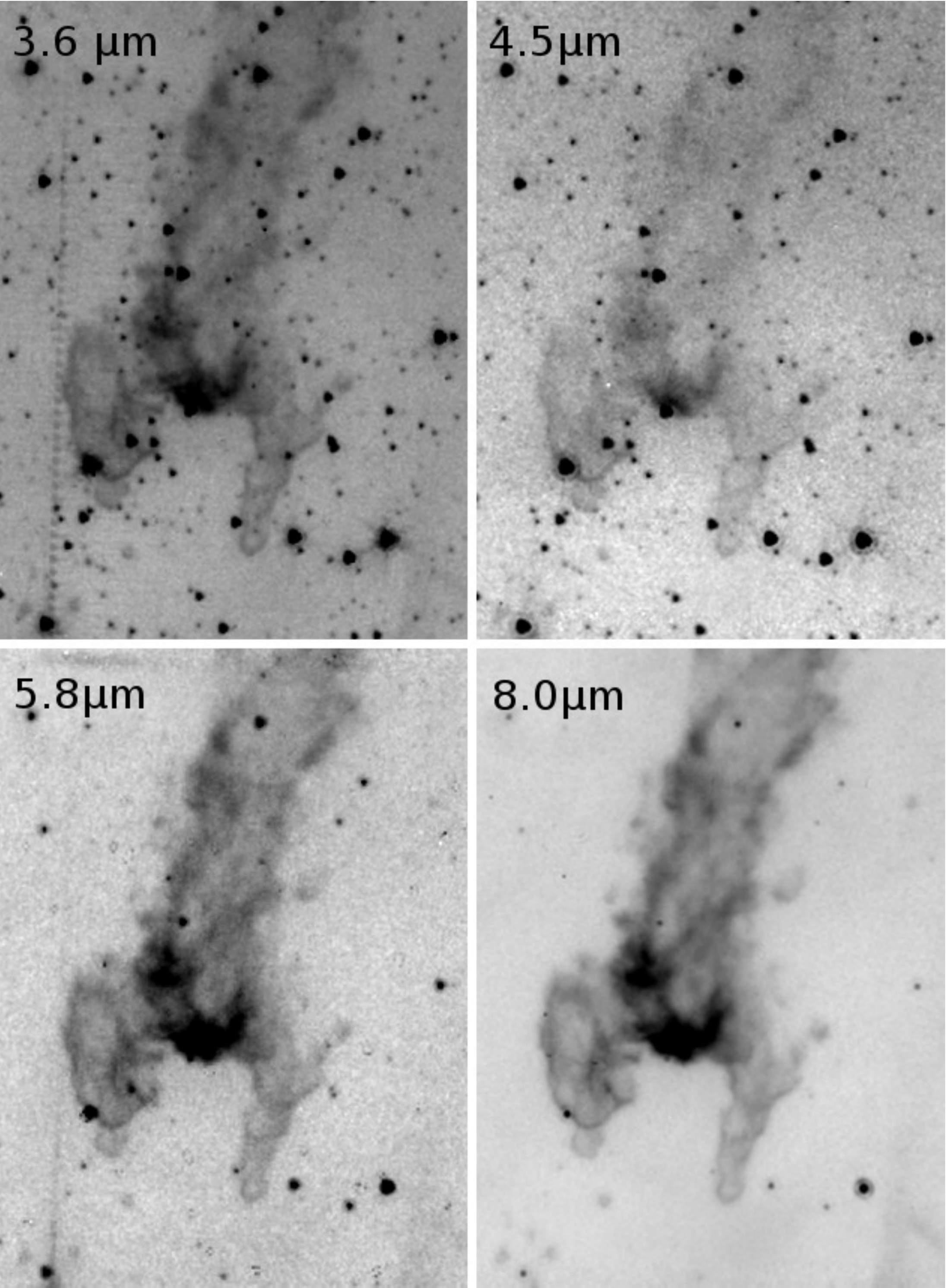}
    \caption{The Wrench in the \textit{Spitzer} IRAC bands. Squareroot scaling has been used. Gray-scale levels are chosen to highlight the structures in the region.}
 \label{fig:wrenchirac}
 \end{figure}

\begin{figure}
 \centering
 \includegraphics [width=8cm]{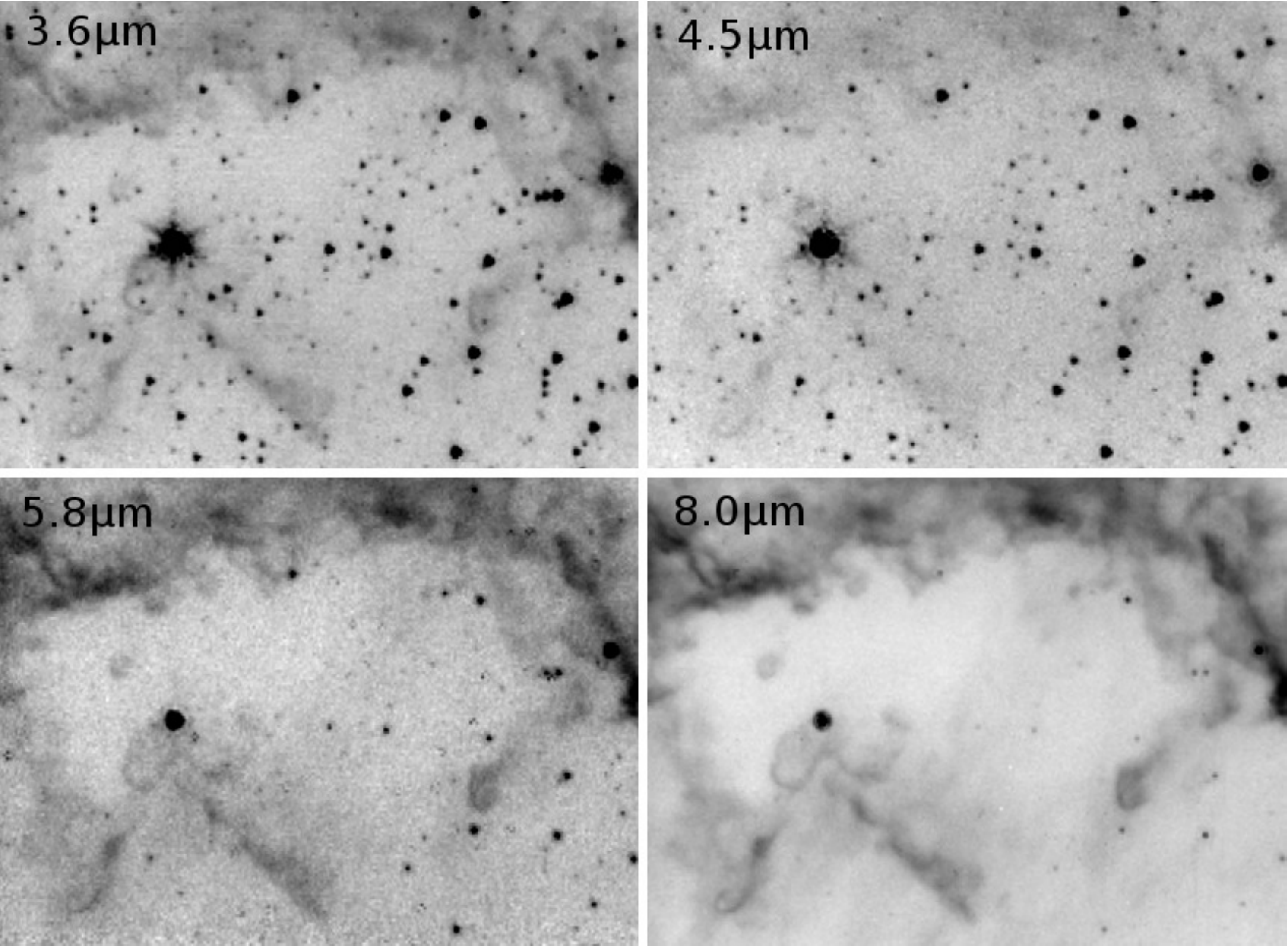}
    \caption{Field 7 in the \textit{Spitzer} IRAC bands. Squareroot scaling has been used. Gray-scale levels are chosen to highlight the structures in the region.}
 \label{fig:f7irac}
 \end{figure}

MIR data allows the study of dust composition and the \Htwo\ 2.12 $\mum$ excitation mechanism. We discuss these large-scale aspects here and the small-scale phenomena of star formation is discussed in Sect. \ref{sect:sf}. IRAC images of the Wrench and F7 are shown as samples in Figs. \ref{fig:wrenchirac} and \ref{fig:f7irac}, respectively.

The surface brightness detected with IRAC traces the \Htwo\ 2.12 $\mum$ surface brightness in the globulettes, trunks, and other structures throughout the entire NIR-imaged area. Especially the bright \Htwo\ rims are bright also in the IRAC images.
We performed aperture photometry on a bright rim in the Wrench and an empty OFF region. We compared the obtained IRAC colors with the shocked \Htwo\ conditions listed in \citet{gutermuthetal08}. The [3.6]-[4.5] and [4.5]-[5.8] colors of the rim are 0.26 and 2.10, respectively, and we estimate that the \Htwo\ 2.12 $\mum$ emission in the rims is not coming from shocked \Htwo. Taking into account a possible $\sim$10\% error in the fluxes, the resulting colors will still indicate that the emission in the rims is not due to shocked \Htwo.

The surface brightness peaks coincide in all IRAC channels. However, the appearance of the 4.5 $\mum$ image differs from the other channels. The surface brightness and small-scale structure of the rims and trunks appear more diffuse and fainter than in the other IRAC channels. For example, in Fig. \ref{fig:f7irac} the edge of the dust shell and the globulette RN 35 are faint at 4.5 $\mum$. The faintness of the 4.5 $\mum$ features can be explained if a large part of the surface brightness is emission from polycyclic aromatic hydrocarbon (PAH) particles. They have strong emission lines in the IRAC bands other than channel 2 \citep{flageyetal06}. However, NIR and MIR emission lines of \Htwo\ also lie within all IRAC filters \citep{blackvandishoeck87, smithrosen05}. Very small grains (VSGs) emit at mid-infrared and are expected to be seen only in the 24 $\mum$ image.

Generally, PAH particles are destroyed in \ion{H}{ii} regions by the strong UV radiation but they can survive behind dense rims where they trail behind the \Htwo\ rim \citep[e.g.][]{kassisetal06}. The density in the globulettes is high already at the outer edge \citepalias{gahmetal13}, indicating that PAHs should also survive close to the surface.
The \textit{Spitzer} resolution is not sufficient to resolve the \Htwo\ and the PAH emitting rim regions from each other at the distance of the Rosette Nebula, but the IRAC images suggest that the globulettes, trunks and the dust shell do contain PAHs. 
Because the 4.5 $\mum$ filter does not contain strong PAH lines, the features in that image therefore have a lower intensity than in the other IRAC bands.

\begin{figure}
 \centering
 \includegraphics [width=8.5cm]{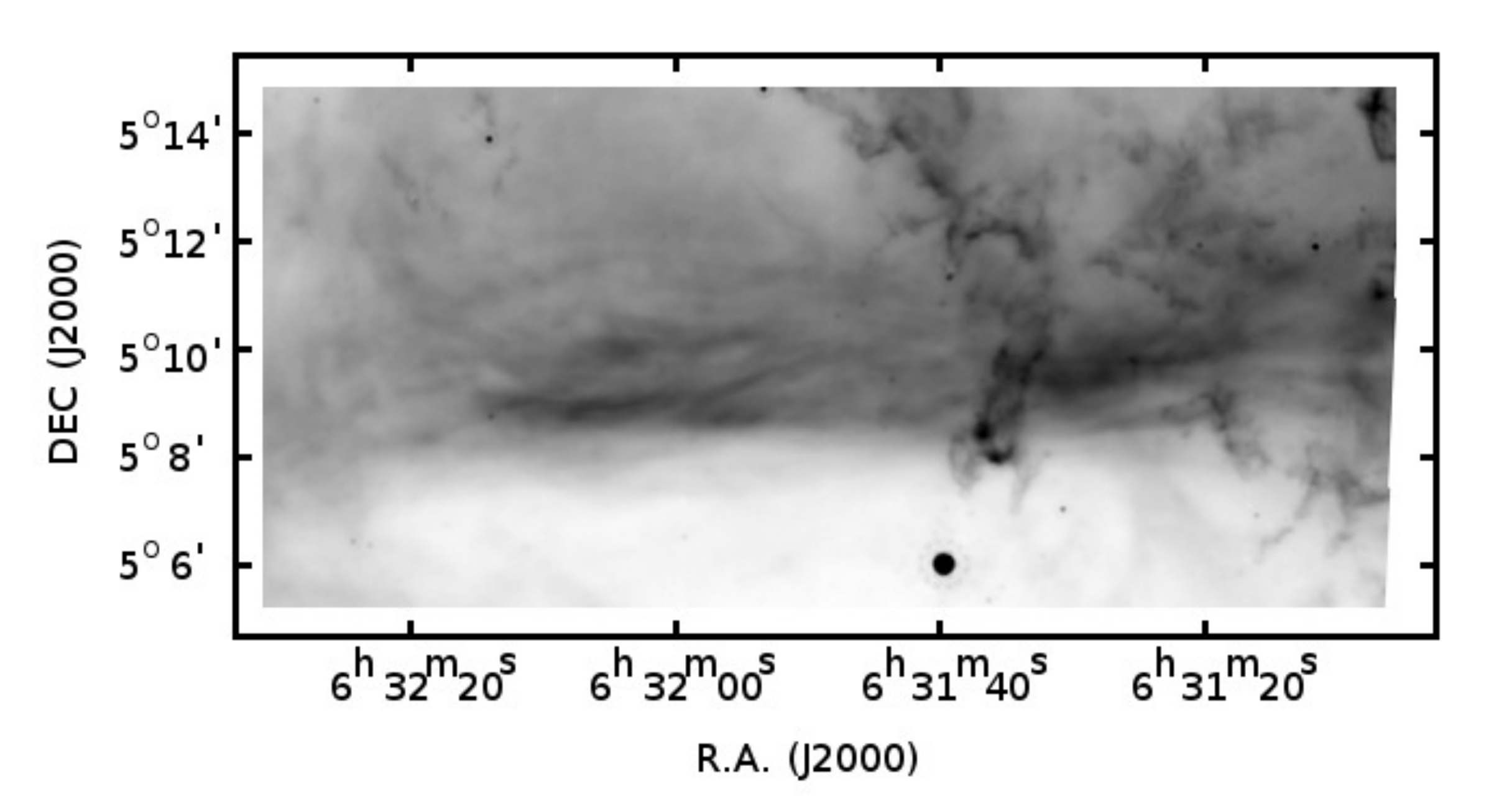}
    \caption{\textit{Spitzer} MIPS 24 $\mum$ image covering fields 8, 14, 15, and 19. Squareroot scaling has been used. Gray-scale levels are chosen to highlight the structures in the region.}
 \label{fig:mips1bw}
 \end{figure}

The 24 $\mum$ MIPS image is shown in Fig. \ref{fig:mips1bw}. Bright rims and extended surface brightness is seen in the trunks, shell, and globulettes.
These bright regions trail behind the NIR-bright rims. The elephant trunks can be seen clearly and especially the Handle of the Wrench is relatively much brighter at 24 $\mum$ than in the \Htwo\ 2.12 $\mum$ image. Fig. \ref{fig:mips1bw} has one distinct feature not seen in the IRAC images except for a faint trace in the 4.5 $\mum$ image. A large extended bright region crosses the 24 $\mum$ image perpendicular to the Wrench starting from F8, crossing the Handle and extending all the way to south of F19. It appears to be a part of the remote side of the nebula and is therefore a background feature. This surface brightness could be explained with thermal emission from VSGs.
We computed the IRAC magnitudes for a small area of the remote Rosette Nebula surface west of the Wrench, where the surface brightness of the background is high at 4.5 $\mum$. 
The resulting colors are [3.6]-[4.5]=1.24 and [4.5]-[5.8]=0.12 which satisfy the conditions \citet{gutermuthetal08} gives for shocked \Htwo. Including a 10\% error in the flux, the conditions are still mostly satisfied, but the [3.6]-[4.5] color will fall below the 1.05 limit (1.02).

\subsection{Herschel imaging: globulette densities}

\begin{figure}
 \centering
 \includegraphics [width=8.5cm]{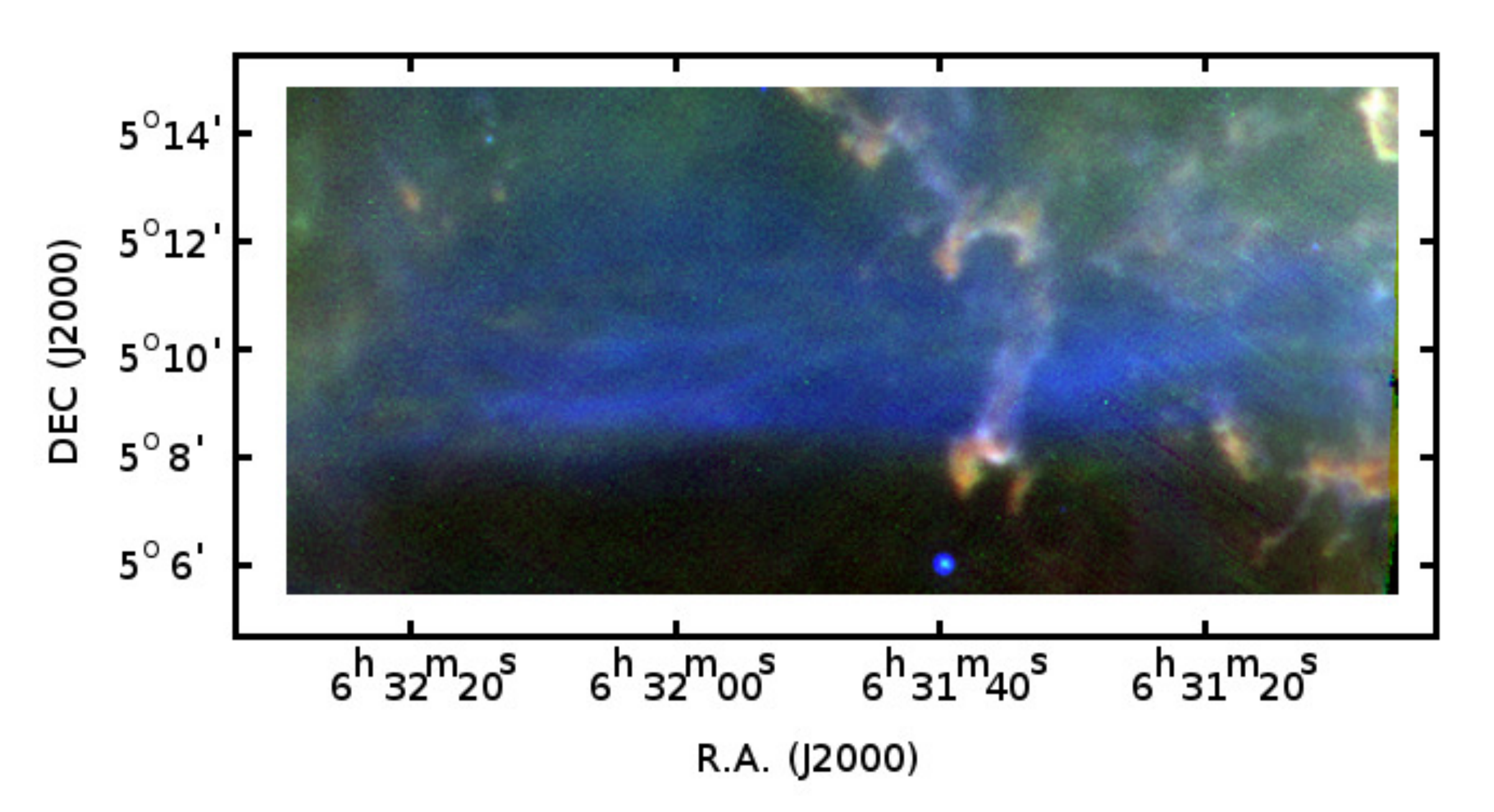}
    \caption{Color-coded MIPS and PACS image covering the SOFI fields except F7. PACS 160 $\mum$, 70 $\mum$, and MIPS 24 $\mum$ are coded in red, green, and blue, respectively. The largest globulettes in F19 can be seen in the upper left corner.}
 \label{fig:pacsmips}
 \end{figure}

Observations of the dust emission in FIR provide further insight on the properties of the Rosette dust shell and the globulettes. The PACS 70 and 160 $\mum$ bands trace thermal emission from small and large grains, respectively. The Herschel PACS images, which cover all but F7 of the SOFI images are shown color-coded together with the Spitzer MIPS 24 $\mum$ image in Fig. \ref{fig:pacsmips}. Gray-scale PACS 70 and 160 $\mum$ images are in Figs. \ref{fig:pacs70neg} and \ref{fig:pacs160neg}, respectively.

The Rosette Nebula dust shell structure is clearly discernible in the PACS images. The rims are bright but the surface brightness extends deep into the shell and trunks behind the rims. The jaws of the Claw, the Wrench and the shell in F8 are the brightest objects in the region. This is in contrast to the 24 $\mum$ image where these features are relatively faint.
The trunks of the tailed globulettes RN 40 and 47 in F8 are seen as well as is a faint trace of the String. The notable difference between the PACS 70 and 160 $\mum$ images is that the 160 $\mum$ surface brightness peaks right behind the 70 $\mum$ features. The 160 $\mum$ surface brightness features correspond well with the features seen in absorption in  P$\beta$. 
In particular, the globulettes RN 95, 114, 122, and 129 are seen both in the  70 $\mum$  and 160  $\mum$ images. The mere detection of these globulettes in the FIR supports the conclusion that some of the globulettes have dense cores.
The small-sized globulette RN 88, which also was seen in absorption in P$\beta$, cannot be seen. This is possibly because of its size which is smaller than the HPBW of the PACS 160 $\mum$ image. The elephant trunks also contain denser clumps.

 \subsection{Recent and on-going star formation}
 \label{sect:sf}

The SOFI and the \textit{Spitzer} archival images of the SOFI regions were searched for signs of recent or on-going star formation. Previous searches for NIR signs of star formation in the Rosette Nebula either did not cover the SOFI fields \citep{balogetal07, ybarraetal13} or located no NIR excess objects in the fields \citep{romanzunigaetal08}. \citet{romanzunigaetal08} found one star with no \J-band detection and with a large \H$-$\Ks\ color that coincides with the eastern tip of the Claw. The \citet{romanzunigaetal08} FLAMINGOS study was however not as deep as this SOFI survey.

\subsubsection{The Wrench}

The surface brightness on the trunk surface in the Cove is seen in all the SOFI NIR images, in the \Htwo\ 2.090 continuum image and in the Br$\gamma$ image. The surface brightness is therefore mostly due to scattered light.
The star \object{2MASS J06313623+0507501} is located just south of the bright rim and is a likely candidate for the origin of the scattered light. This location of the star just off the elephant trunk suggests the star was most likely formed inside the trunk and then exposed due to the UV field evaporating the trunk material.
The \textit{Spitzer} Enhanced Imaging Products (EIP) Source List includes it as SSTSL2 J063136.23+050750.0 and gives it IRAC magnitudes of 12.16, 12.15, 11.76, and 11.01 for channels 1 to 4, respectively. At 3.6 $\mum$ the EIP catalog lists the bandfilled magnitude, indicating the source detected in another wavelength. The resulting colors have been plotted in Fig. \ref{fig:ysoplot} along with the typical YSO distributions of \citet{poultonetal08}. The star is located at the edge of the region where the stellar energy distributions (SED) can be modeled with a stellar photosphere.
We obtained a spectral class of A2 V for this star from spectrograms taken for us through the SMARTS consortium, Cerro Tololo. This along with the SOFI NIR colors (\J$-$\H=0.27, \H$-$\Ks=0.17) suggests a visual extinction of $\sim$2.5\umag. A2 stars have a typical mass of $\sim$2 $\Msun$ and a ZAMS age of about $8\times10^{6}$ yr.

\subsubsection{The Claw}

\begin{figure}
 \centering
 \includegraphics[width=8cm]{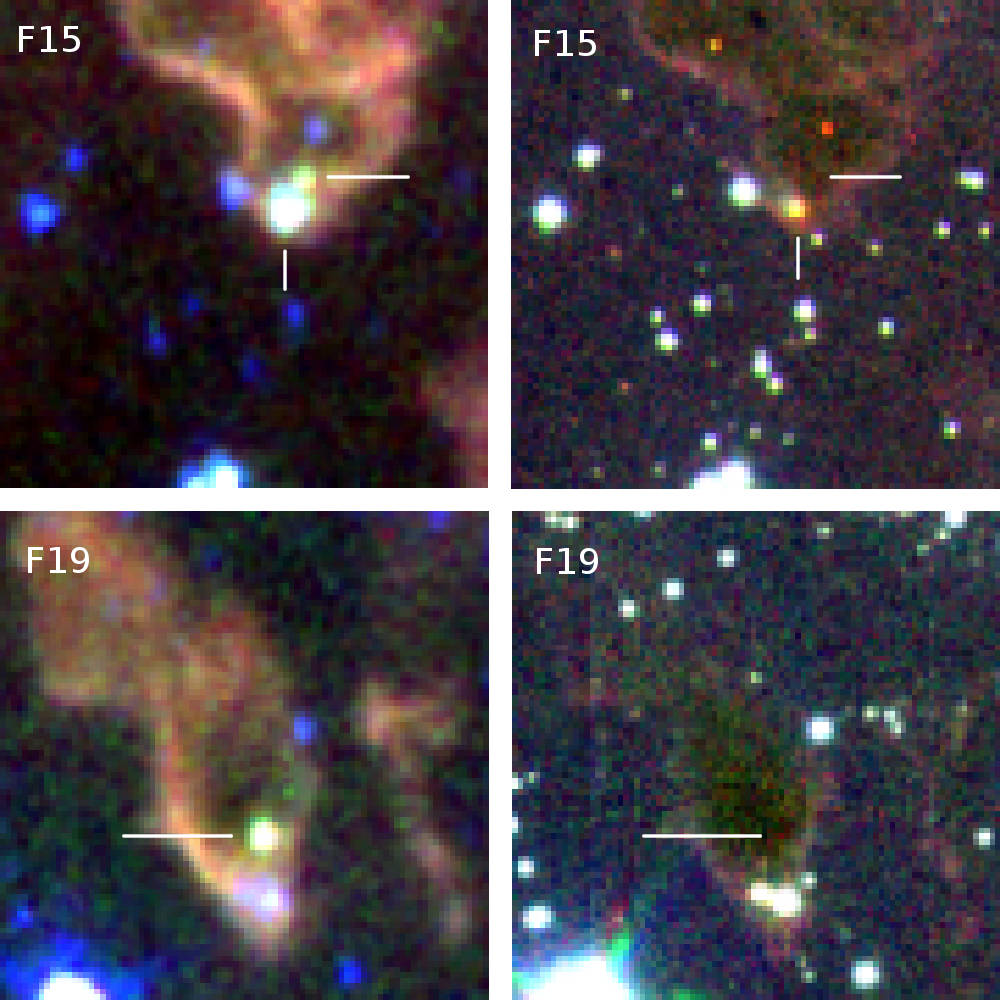}
   \caption{False-color images of the eastern jaw of the Claw (upper panels) and of globulette RN 129 (lower panels). The lines indicate the locations of YSOs. Left: IRAC bands 8.0, 5.8, and 3.6 $\mum$ coded in red, green, and blue, respectively. Right: SOFI \Ks, \H, and \J\ images coded in red, green, and blue, respectively. All images are 53\arcsec$\times$53\arcsec.}
 \label{fig:ysos}
 \end{figure}

The tip of the eastern jaw of the Claw is an active region with two IR excess objects (Fig. \ref{fig:ysos}, upper panels). The brighter star is detected on the bright \Htwo\ rim of the jaw in all SOFI and IRAC images. A second, fainter object is visible only in the IRAC images and it lies 3.6\arcsec\ north-northwest of the first toward the core of the jaw. The \textit{Spitzer} 24 $\mum$ image shows an object at the jaw tip, but because of the low spatial resolution this image could trace either one or both of the IR excess objects. The red object in the middle of the core in the SOFI image in Fig. \ref{fig:ysos} is a highly reddened background star discussed in Sect. \ref{sect:globav}.

\begin{figure}
 \centering
 \includegraphics[width=8cm]{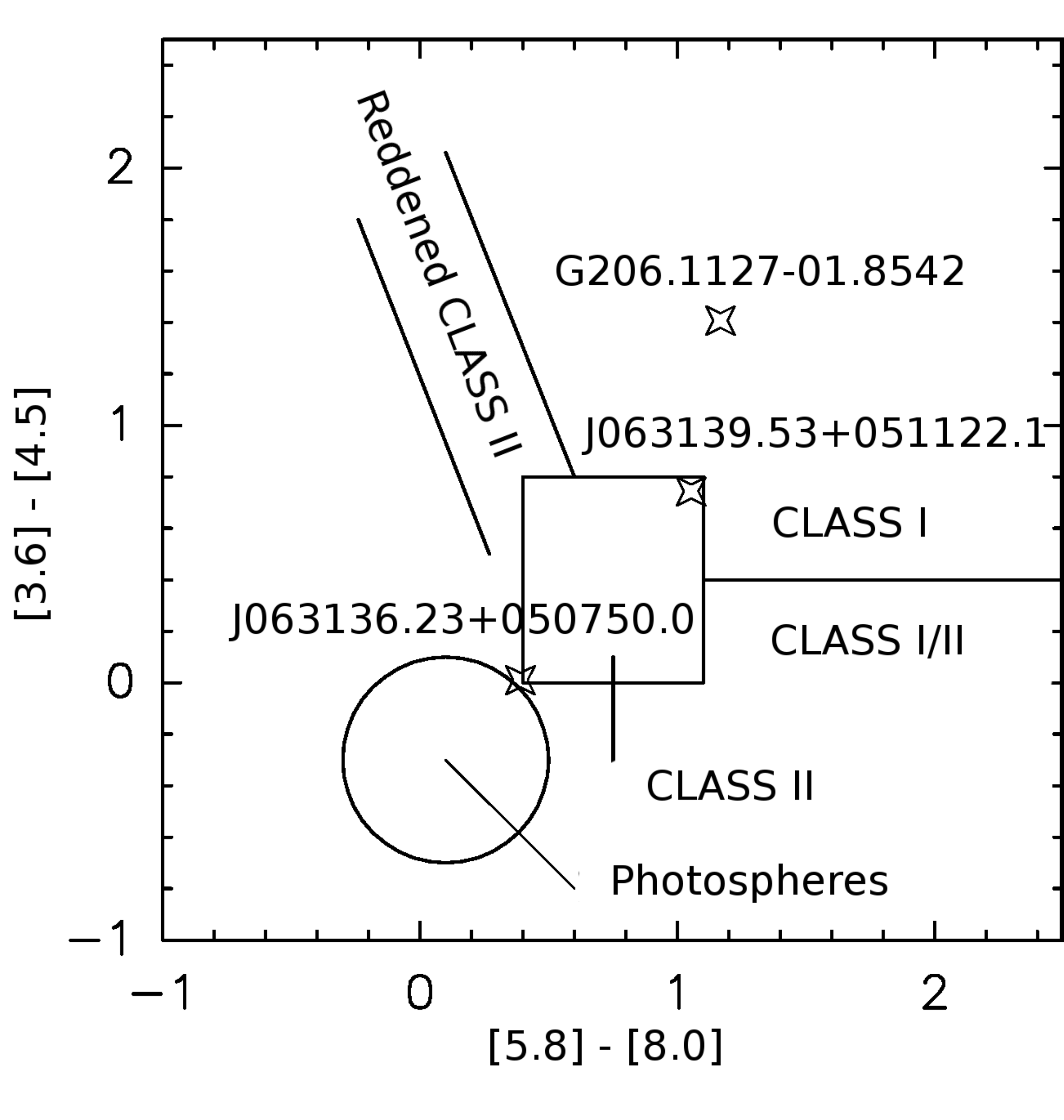}
   \caption{The IRAC colors of the A2 star and the YSO candidates. Stars mark the objects and the lines are the YSO Class conditions of \citet{poultonetal08}. Class III stars are located inside the circle and Class II sources inside the box. The upper left-hand corner lines describe the reddened Class II and the horizontal line Class I/II.}
 \label{fig:ysoplot}
 \end{figure}

The brighter object on the rim has \jhks\ colors \J$-$\H=2.37 and \H$-$\Ks=1.17 in our initial photometry catalog. It was not discovered by \citetalias{gahmetal13} because it was not included in the final SOFI photometry catalog. It is listed in the Spitzer EIP catalog as \object{SSTSL2 J063139.53+051122.1}.
However, the available EIP catalog flux values in the 4.5 and 8.0 $\mum$ bands are bandfilled fluxes. With aperture photometry we obtained IRAC magnitudes 13.32, 12.58, 11.34, and 10.288 for the brighter component.
The Spitzer EIP catalog also gives [24] = 6.72\umag\ and based on the IRAC images, most of that flux likely belongs to the brighter component. These \textit{Spitzer} magnitudes put the brighter object just inside the Class II box in the [3.6]-[4.5] vs. [4.5]-[5.8] diagram of \citep{poultonetal08} (see Fig. \ref{fig:ysoplot}).
The colors also satisfy the protostar condition in \citet{gutermuthetal08}, but if the unknown foreground extinction toward the object is sufficiently high ($\sim20$\umag), it could be even a Class II source. A 10\% error in the fluxes introduces a $\pm$0.2\umag\ error to the colors. We consider the brighter object to be a young stellar object (YSO), possibly of Class I/II, and the IR images confirm that the YSO is located inside the elephant trunk.

We consider the fainter IR excess object also to be a YSO candidate. It does not have any photometric data but in the images it is a faint detection in IRAC band 1 and a clear detection in the longer IRAC wavelengths.

\subsubsection{Globulette RN 129}
\label{sect:protostar129}

IRAC and SOFI \jhks\ images of globulette RN 129 are shown in Fig. \ref{fig:ysos} (lower panels). The IRAC images show an IR excess object 6.8\arcsec\ north of the bluish star located on the rim. It is seen also in the 24 $\mum$ image but not in the \jhks\ NIR images. The \textit{Spitzer} GLIMPSE360 Catalog object \object{SSTGLMC G206.1127-01.8542} corresponds to this object. The catalog only provides the magnitude at 4.5 $\mum$ (14.36\umag). MOPEX photometry with a 3.6\arcsec\ aperture radius gives IRAC magnitudes 15.01, 13.61, 12.21, and 11.04.
The resulting colors satisfy the protostar condition of \citet{gutermuthetal08} even when the 0.2\umag\ color errors are included. In the \citet{poultonetal08} IRAC color-color diagram in Fig. \ref{fig:ysoplot}, the star is clearly placed in the Class I section.

\subsection{Fluorescent \Htwo}

We have observed bright \Htwo\ 1--0 S(1) line emission at 2.12 $\mum$ in the trunks and globulette rims, but based on the IRAC colors, it does not appear to be due to shocked \Htwo\ (see Sect. \ref{sect:spitzer}). Therefore, an alternative emission mechanism has to be responsible. We suggest that fluorescent \Htwo\ could explain the bright \Htwo\ rims.

Fluorescence takes place as a result of FUV photons from 912 to 1108 \AA\ exciting the \Htwo\ molecules. The molecules then partly decay via fluorescence and are observed in NIR. The probabilities for each transition in the cascade to lower levels can be computed using different models for e.g. cloud n$_{H}$ densities and UV field strengths. The resulting spectrum is very sensitive to these two variables and the (N)IR lines  can be used to determine the conditions in the \Htwo. \citet{blackvandishoeck87} computed line strengths for fluorescent \Htwo\ emission in several models. The different models show that the \Htwo\ 1--0 S(1)/ 2--1 S(1) ratio is fairly constant, $\sim$1.7-2.

In the globulettes, the \Htwo\ emission is seen close behind the NOT H$\alpha$ front, indicating that hydrogen is in molecular form just beneath the globulette surface \citepalias{gahmetal13}. The \Htwo\ emission comes from a thin skin on the surface of the globulettes.  However, in the smaller globulettes the UV photons can penetrate the globulette, depending on its density and the strength of the UV field.

The observed \Htwo\ and \Ks\ fluxes in globulettes (Tables \ref{table:coresb} and \ref{table:rimsb}) suggest that about a third of the \Ks\ band flux is due to the \Htwo\ 1--0 S(1) 2.12 $\mum$ emission. Other lines in the \Ks\ filter are needed to explain the remaining two thirds of the \Ks\ flux. If fluorescence is causing the \Htwo\ emission, then $\sim$ 20\% of the total line emission in the \Ks\ band would be due to the 2--1 S(1) line and about half of the emission would come from other \Htwo\ lines. In the case of thermal emission, the 1--0/total ratio would be higher than in the fluorescent case as the higher emission lines would be weaker \citep{blackvandishoeck87}.

\subsubsection{Comparison with Eagle Nebula and Horsehead Nebula}

The Eagle Nebula (M16) contains the well-known elephant trunks imaged e.g. by \citet{hesteretal96}. These elephant trunks are exposed to UV emission from a cluster of OB stars which is similar to the situation in the Rosette Nebula. \citet{allenetal99} detected fluorescent \Htwo\ in the PDRs in the heads of the elephant trunks. Surface brightness studies of images in the \Htwo\ 1--0 S(1) and \Htwo\ 2--1 S(1) lines showed regions of both pure fluorescence and fluorescence boosted by collisional excitation. Their peak line intensities of the \Htwo\ 1--0 S(1) line is of the same order as the surface brightness in the Rosette Nebula globulette rims and the line emission traces the surface of the M16 trunks. However, they detected strong Br$\gamma$ emission which is almost absent in the Rosette Nebula. NIR observations of the elephant trunks by \citet{sugitanietal02} showed that the bulk of the trunks is less dense than the core that shields the trunks from ionizing radiation coming from the central cluster in the Eagle Nebula. Similar morphology is seen in the Rosette Nebula with the dense cores shielding the trunks that form tails behind the cores.

The Horsehead Nebula has been observed with \textit{Spitzer}, and the morphology of the PDR is very similar in all IRAC channels. The 4.5 $\mum$ emission in the PDR behaves similarly to the Rosette Nebula globulettes even though the effect is not as distinct in the Horsehead Nebula. \citet{habartetal05} detected a very narrow (5\arcsec) fluorescent \Htwo\ 1--0 S(1) filament which is viewed edge-on. At the distance of the Rosette Nebula this filament would be $\sim$1.5\arcsec\ wide which corresponds well with the width of the bright rims in the Rosette Nebula globulettes. \citet{compiegneetal07} used the IRS spectrograph (5.2-38 $\mum$) onboard \textit{Spitzer} and detected PAHs and rotational \Htwo\ 0--0  S(4) to S(0) lines at the PDR, and some of the rotational \Htwo\ and PAH lines further inside the cloud.

The physical situation in the Rosette Nebula is probably very similar to the Eagle and Horsehead Nebulae. The results from those suggest that the globulettes, trunks, and shells in the Rosette Nebula contain both \Htwo\ and PAHs even though their detailed distribution in the Rosette Nebula cannot be probed with the data presented here.


\section{Summary and Conclusions} \label{sect:conclusions}

We have used new NIR broad- and narrow-band images to study the shells, elephant trunks, and globulettes in the NW quadrant of the Rosette Nebula. We catalog the \Htwo\ 2.12 $\mum$ and P$\beta$ detections in the globulettes. We compute the surface brightness for the rims and cores of some globulettes. A visual extinction map and globulette mass estimates have been made. Additionally, we used archive data from the \textit{Spitzer} and \textit{Herschel} satellites. These are used to search for young pre-main-sequence stars, estimate PAH contribution to the surface brightness,  map the IR emission from the elephant trunks and other shell features, and get a general view of the Rosette Nebula shell.

\begin{itemize}
\item The majority of the smallest globulettes discovered in an optical survey \citepalias{gahmetal07} can be detected in NIR via surface brightness in \Htwo\ 1--0 S(1) 2.12 $\mum$ line, but they are too small to be studied in detail. Some of the larger globulettes, trunks, and parts of the shell appear dark against the $P\beta$ background emission. The 160 $\mum$ dust thermal emission corresponds well with these areas, confirming that they have dense cores. The globulette sizes in the NIR and H$\alpha$ images suggest that the density is high already at the outer edge of the globulettes as was also concluded from molecular line data in \citetalias{gahmetal13}. A survey of stars seen behind globulettes show typical number densities to be $n(\Htwo) =$ a few $\times10^{-4}$ cm$^{-3}$. The masses estimated from this indicate that the larger globulettes contain more mass than estimated in the optical survey, but confirms that the globulette masses are subsolar.
\item About one-third of the core surface brightness (0.26$\pm$0.02) and rim brightness (0.30$\pm$0.01) observed in \Ks\ is caused by the \Htwo\ 1--0 S(1) line, and the remaining two-thirds are from other emission lines. These values suggest that fluorescent \Htwo\ excitation and not shock excitation is the cause of observed \Htwo\ 2.12 $\mum$ emission in the bright rims seen in most globulettes and trunks.
\item IRAC observations suggest that the trunks and globulettes contain also PAHs like a typical PDR. In addition to the shell features and trunks seen in the NIR images, the MIPS and PACS images also trace dust emission from the remote side of the Rosette Nebula.
\item Two globulettes/trunks out of the detected 84 contain YSOs. The eastern jaw of the Claw contains two YSOs separated by 3.6\arcsec. The fainter is not detected in the SOFI images, and the brighter is listed in the Spitzer EIP catalog as SSTSLP J063139.53+051122.1. The YSO classes cannot be determined accurately due to the uncertainty in the amount of foreground A$_{V}$. RN 129 contains the star SSTGLMC G206.1127-01.8542 that is not detected in the SOFI images. Its \textit{Spitzer} colors indicate that it is a class 0/I YSO.

\end{itemize}

\subsection{Future work}

High-resolution NIR spectroscopy of the globulette rims is needed to determine which  lines contribute to the bright rims. The $n_{H}$ of the globulettes and the UV field present in the Rosette Nebula could be determined using ratios of the fluorescent \Htwo\ lines. The contribution of PAHs can also be evaluated via spectroscopy. More detailed, better sensitivity and spatial resolution observations of the globulettes and YSOs will give deeper insight on the star or brown dwarf/free-floating planet formation in globulettes.

\begin{acknowledgements}

M.M. acknowledges the support from the Doctoral Programme in Astronomy and Space Physics. We thank Fred Walter for arranging spectroscopic observations on Cerro Tololo. This research has made use of NASA's Astrophysics Data System, and the SIMBAD database, operated at CDS, Strasbourg, France. This work is based in part on observations made with the {\textit Spitzer Space Telescope}, which is operated by the Jet Propulsion Laboratory, California Institute of Technology under a contract with NASA, and with the \textit{Herschel Space Observatory.} \textit{Herschel} is an ESA space observatory with science instruments provided by European-led Principal Investigator consortia and with important participation from NASA.
\end{acknowledgements}

\bibliographystyle{aa}
 \bibliography{RNglobulettes_refarticles.bib}

\listofobjects

\Online

\begin{appendix}

\section{Online figures}
 \label{app:figures}

\begin{figure*}
 \centering
 \includegraphics [width=17cm] {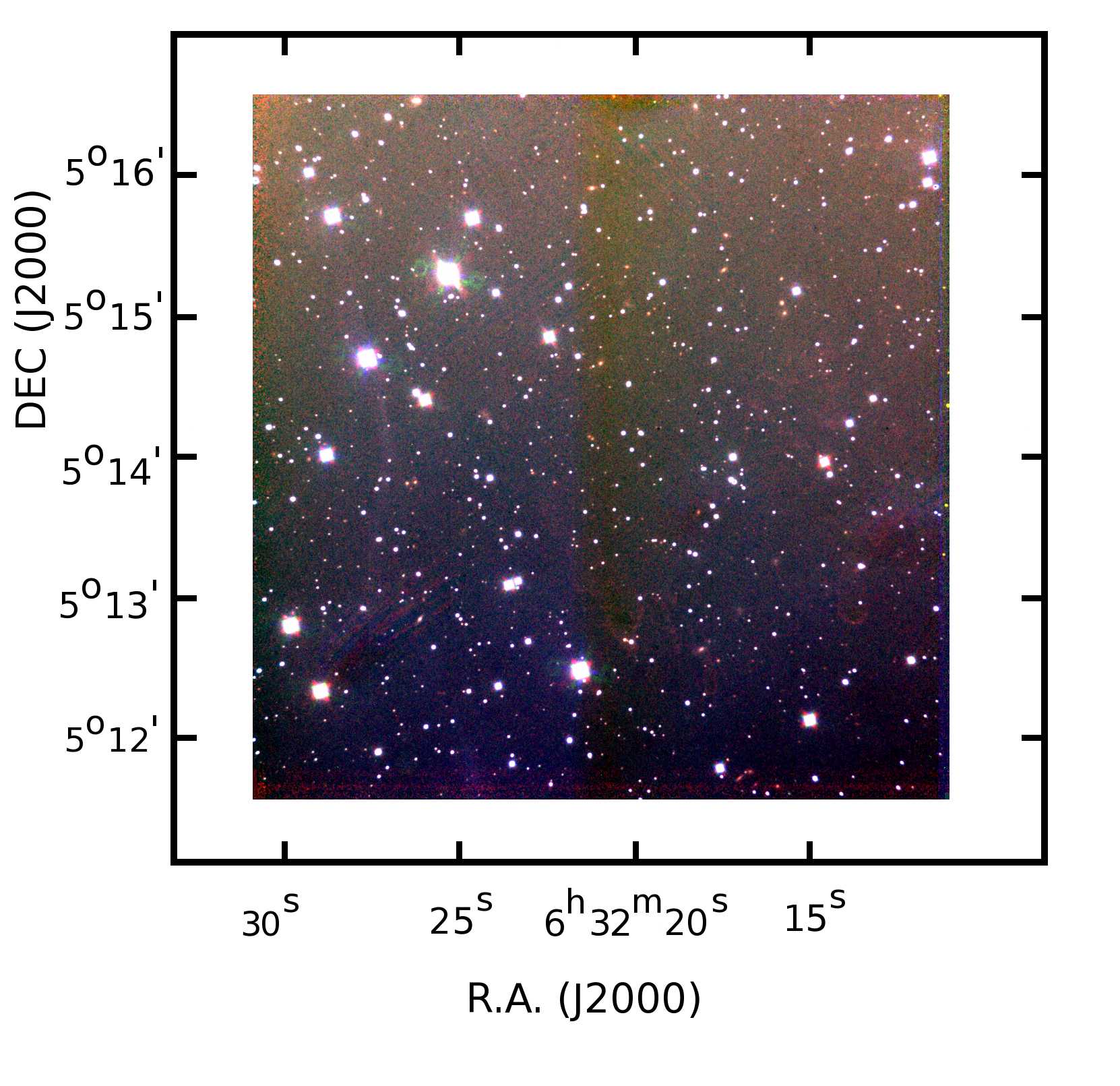}
    \caption{Color-coded SOFI image of F19 obtained in on-off mode. The \J, \H, and \Ks\ bands are coded in blue, green, and red.}
 \label{fig:f19jhk}
 \end{figure*}

\begin{figure*}
 \centering
 \includegraphics [width=17cm] {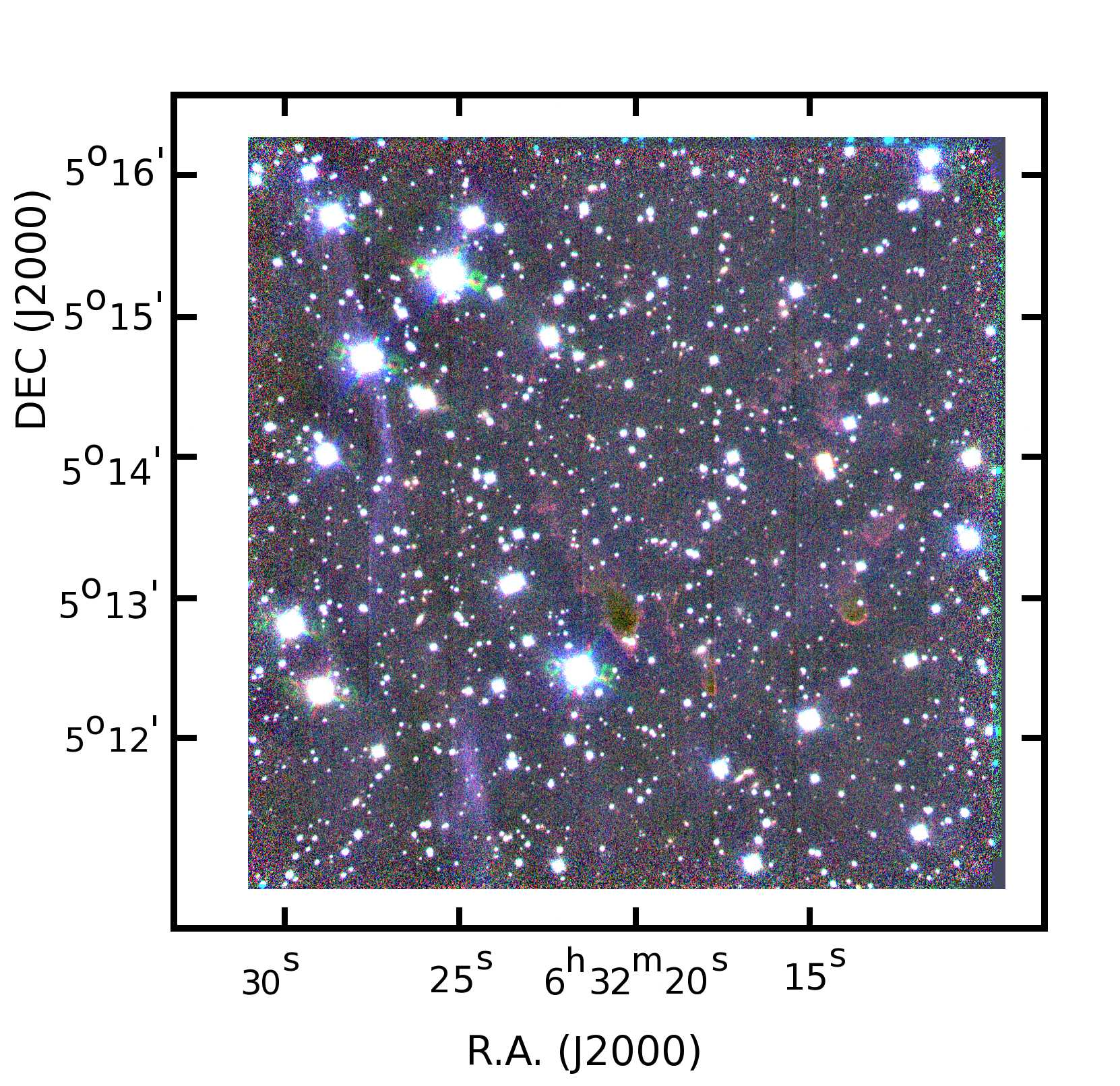}
    \caption{Color-coded SOFI image of F19 obtained in jitter mode. The \Js, \H, and \Ks\ bands are coded in blue, green, and red.}
 \label{fig:f19jhkjitter}
 \end{figure*}

\begin{figure*}
 \centering
 \includegraphics [width=6cm]{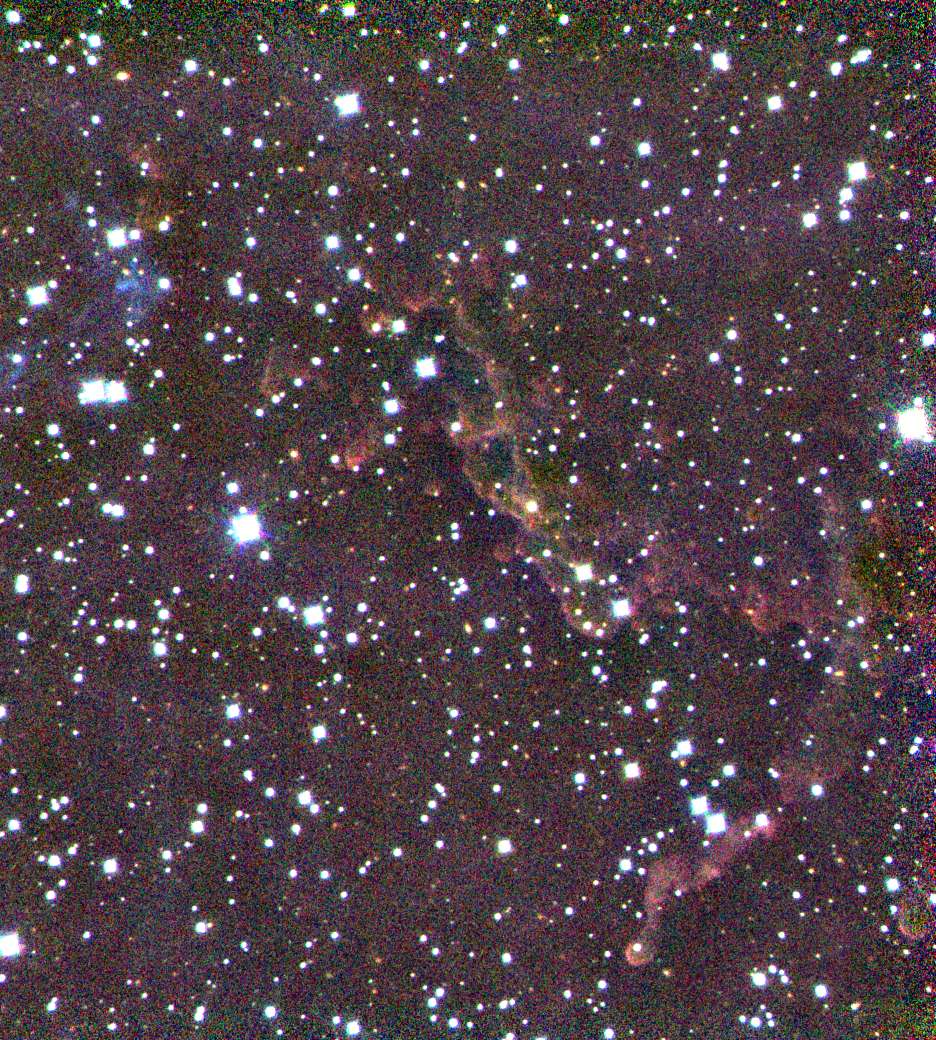}
 \includegraphics [width=6cm]{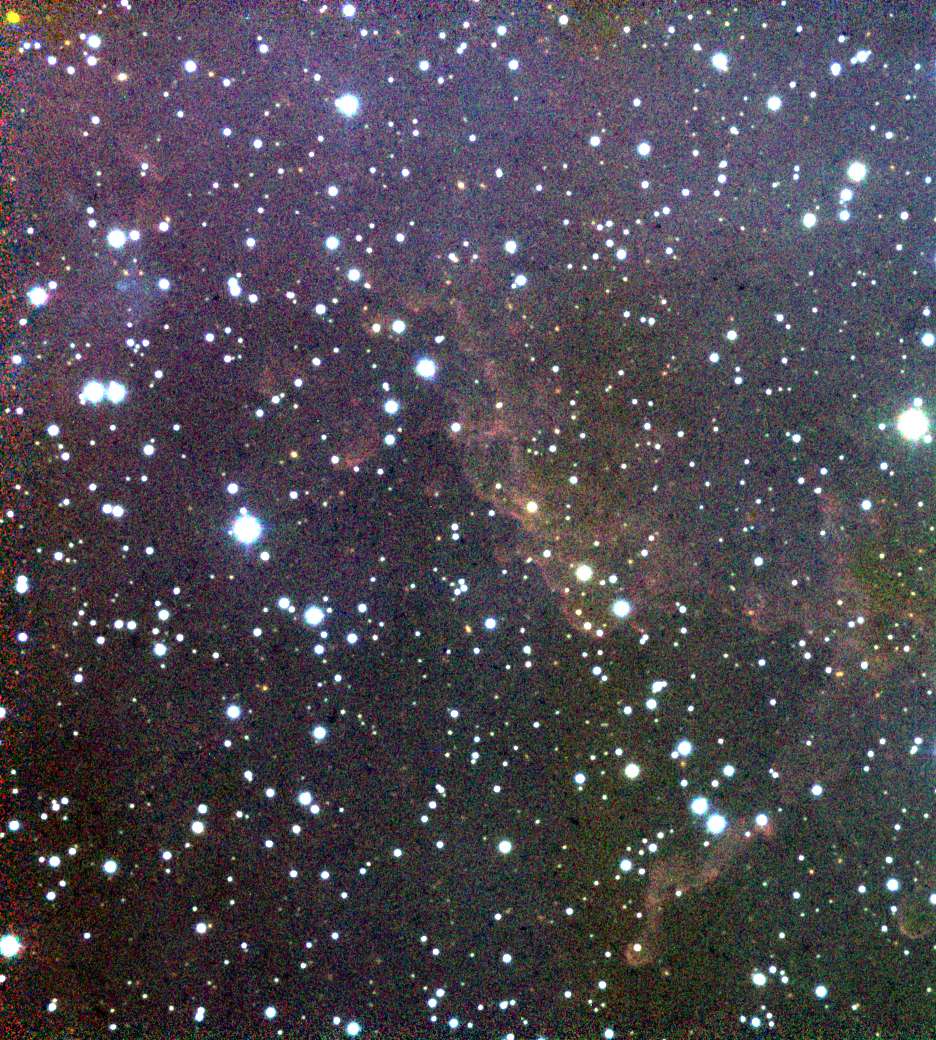}
    \caption{Jittered \jshks\ (left) and on-off \jhks\ (right) image of F8.}
 \label{fig:f8jhkjitter}
 \end{figure*}

\begin{figure*}
 \centering
 \includegraphics [width=14cm]{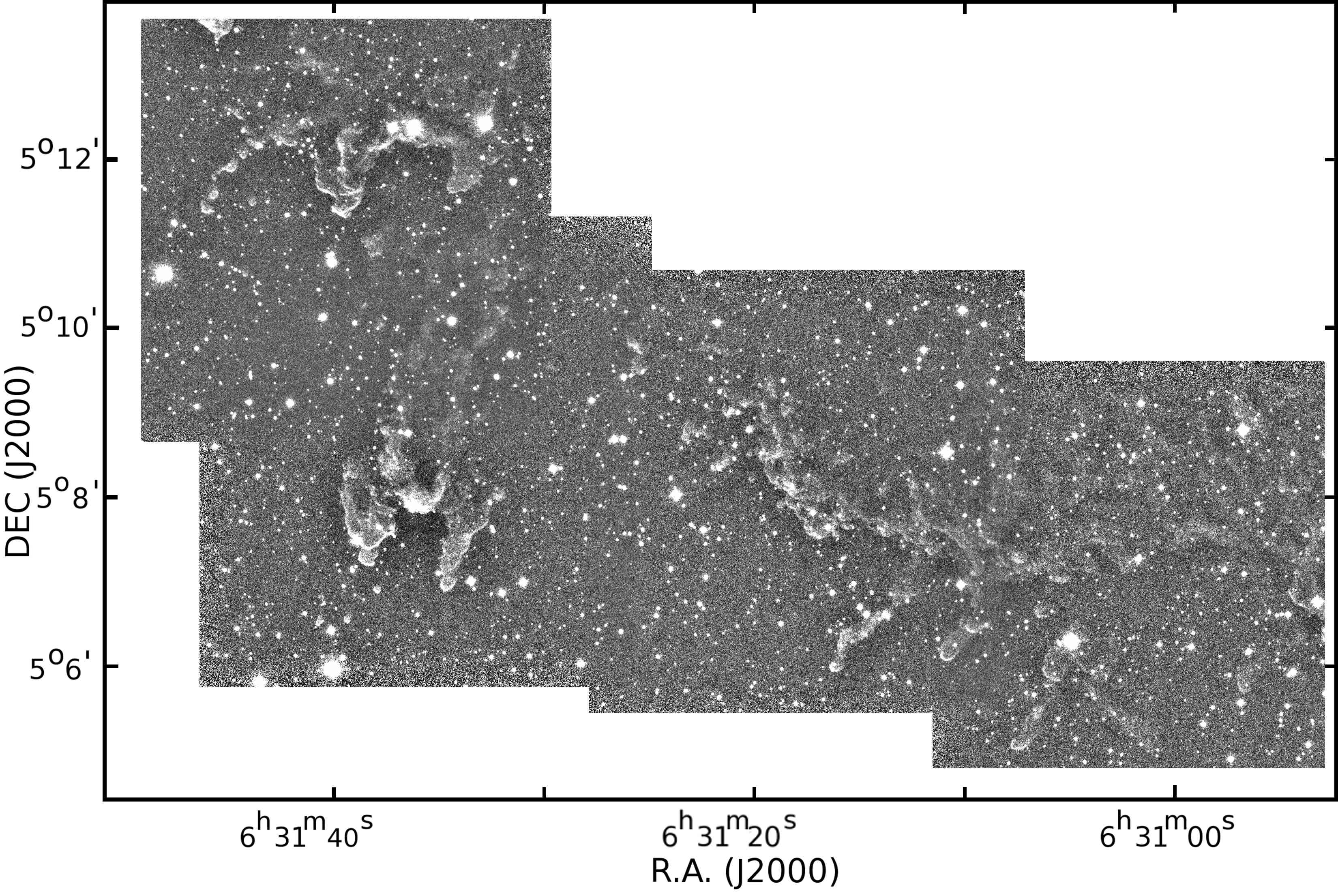}
    \caption{NB 2.124 \Htwo\ S(1) image of fields 7, 8, 14, and 15.}
 \label{fig:h2map}
 \end{figure*}

\begin{figure*}
 \centering
 \includegraphics [width=14cm]{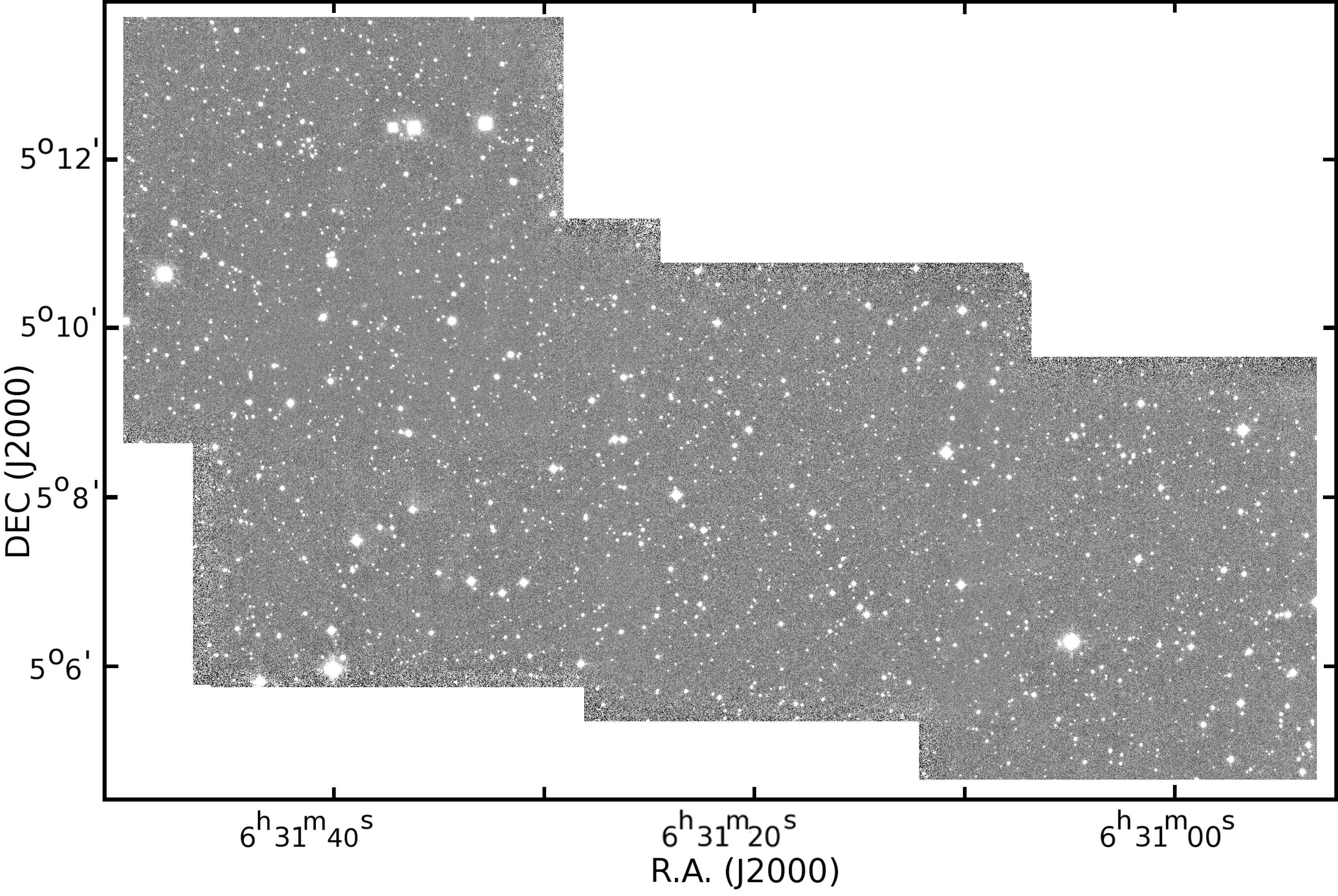}
    \caption{NB 2.090 continuum image of fields 7, 8, 14, and 15.}
 \label{fig:contmap}
 \end{figure*}

\begin{figure*}
 \centering
 \includegraphics [width=14cm]{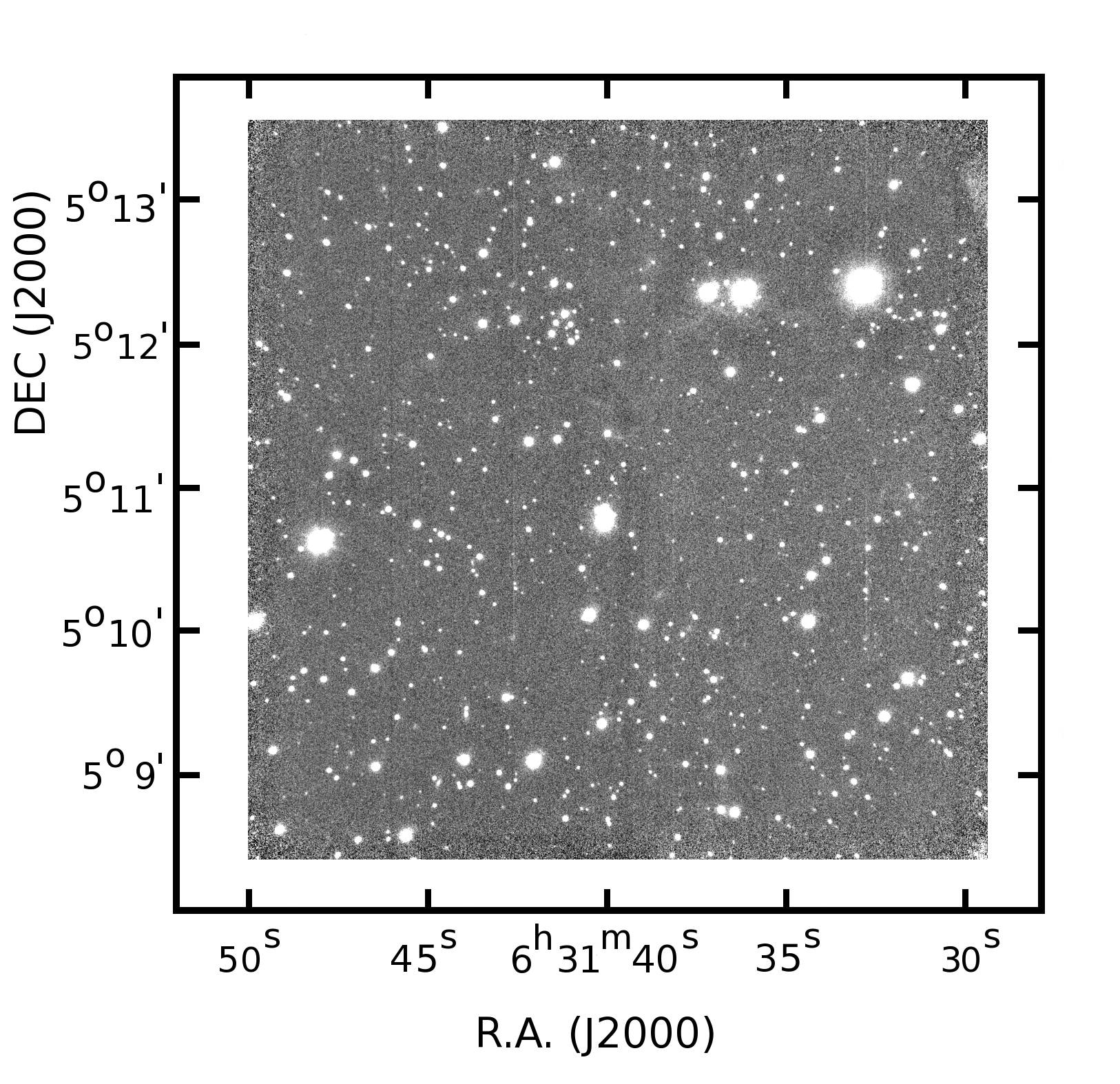}
    \caption{NB 1.257 continuum image of F15.}
 \label{fig:jcontmap}
 \end{figure*}

\begin{figure*}
 \centering
 \includegraphics [width=14cm]{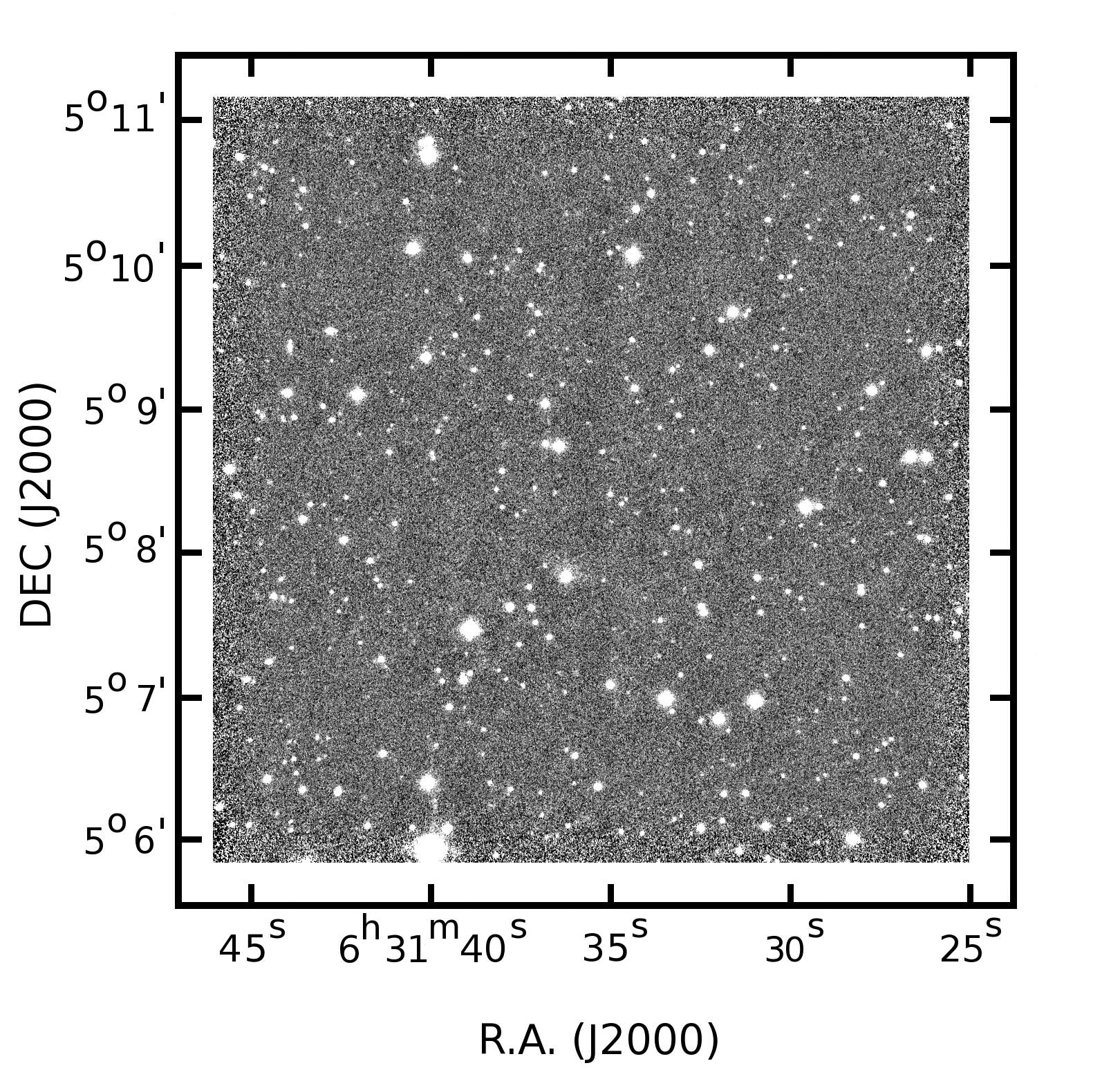}
    \caption{NB 2.167 Br$\gamma$ image of F14.}
 \label{fig:brgmap}
 \end{figure*}

\begin{figure*}
 \centering
 \includegraphics [width=16cm] {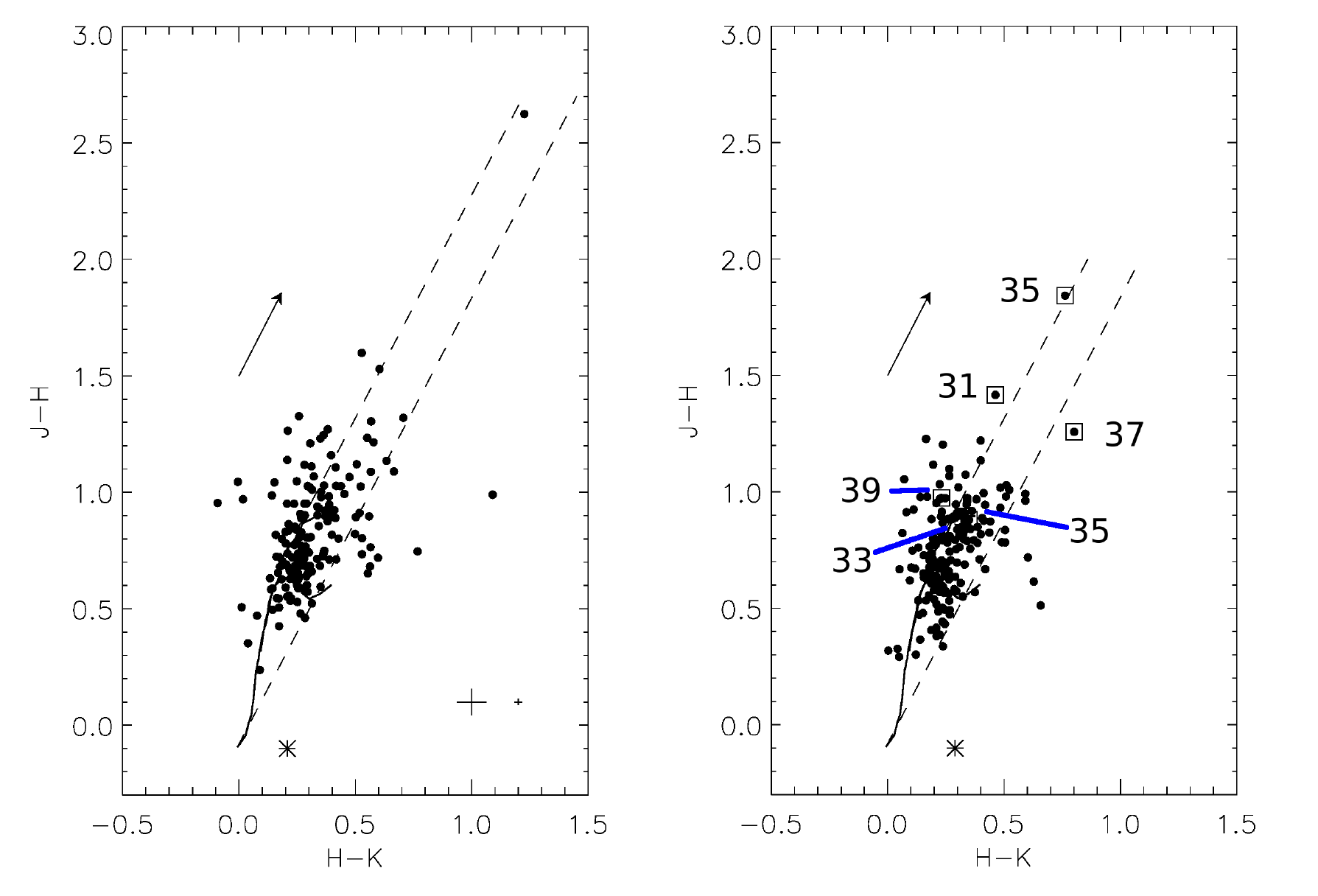}
    \caption{The \J$-$\H, \H$-$Ks\ color-color diagram of the shell (left panel) and background (right panel) objects in F7. Filled circles mark the observed stars and boxes mark the stars behind the globulettes. Asterisks mark stars with only \H$-$\Ks\ colors. The crosses in the bottom right-hand corner indicate the maximum (left) and the average error (right). A reddening vector of 3\umag\ is indicated.}
 \label{fig:jhhkdiagramf7}
 \end{figure*}

\begin{figure*}
 \centering
 \includegraphics [width=16cm] {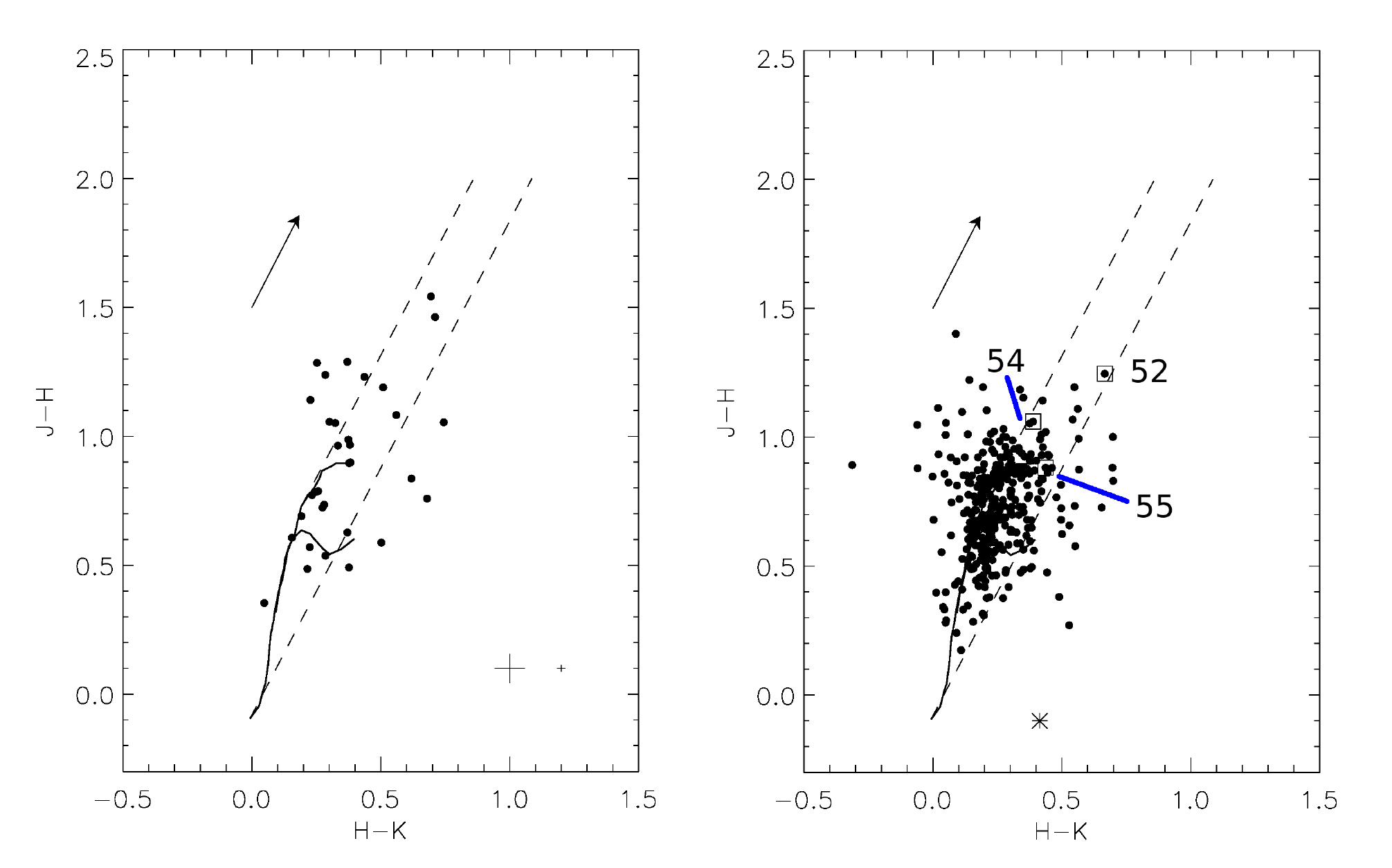}
    \caption{As Fig. \ref{fig:jhhkdiagramf7} of F8.}
 \label{fig:jhhkdiagramf8}
 \end{figure*}

\begin{figure*}
 \centering
 \includegraphics [width=16cm] {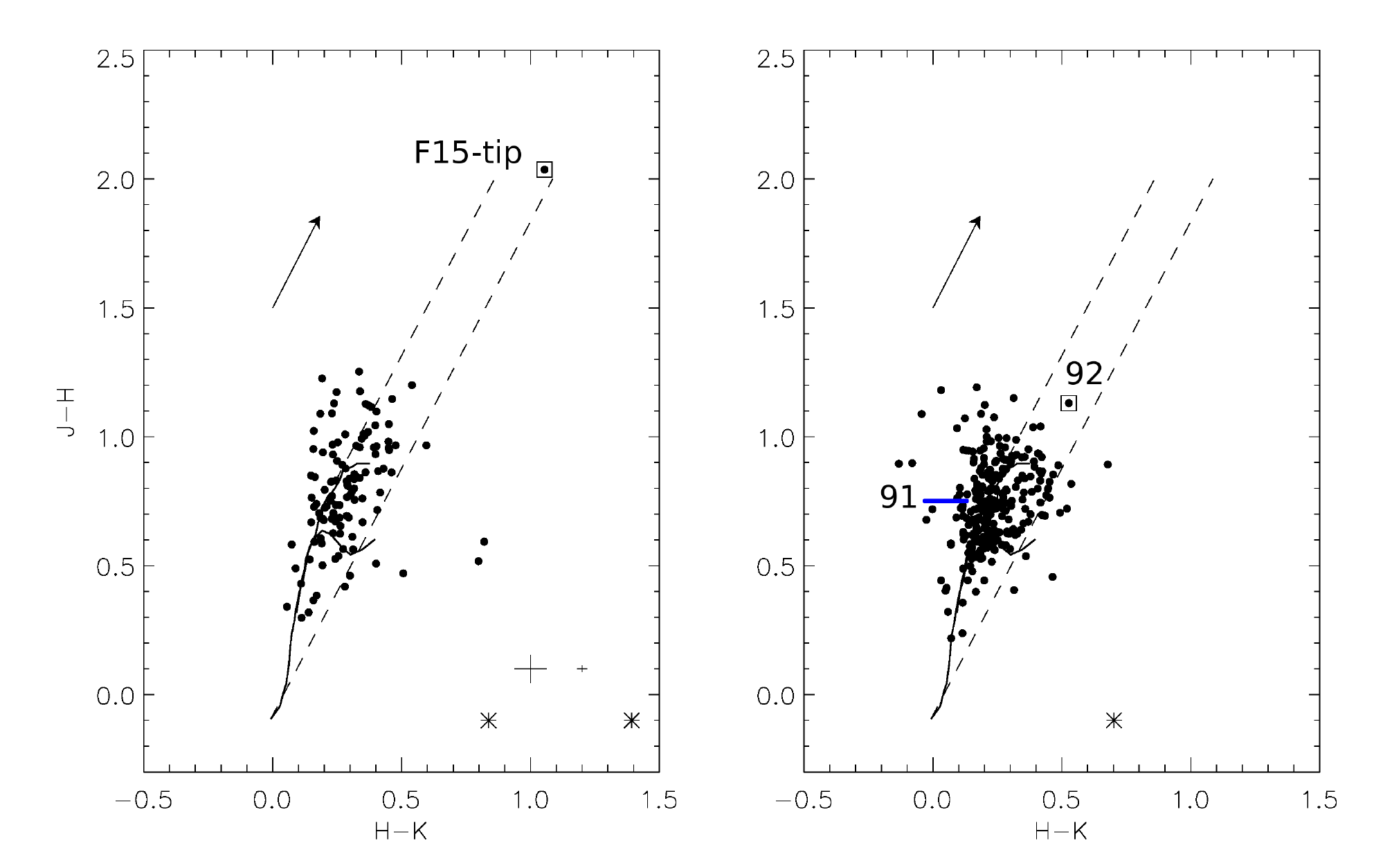}
    \caption{As Fig. \ref{fig:jhhkdiagramf7} of F15. F15-tip is at the northern edge of frame F15.}
 \label{fig:jhhkdiagramf15}
 \end{figure*}

\begin{figure*}
 \centering
 \includegraphics [width=8cm] {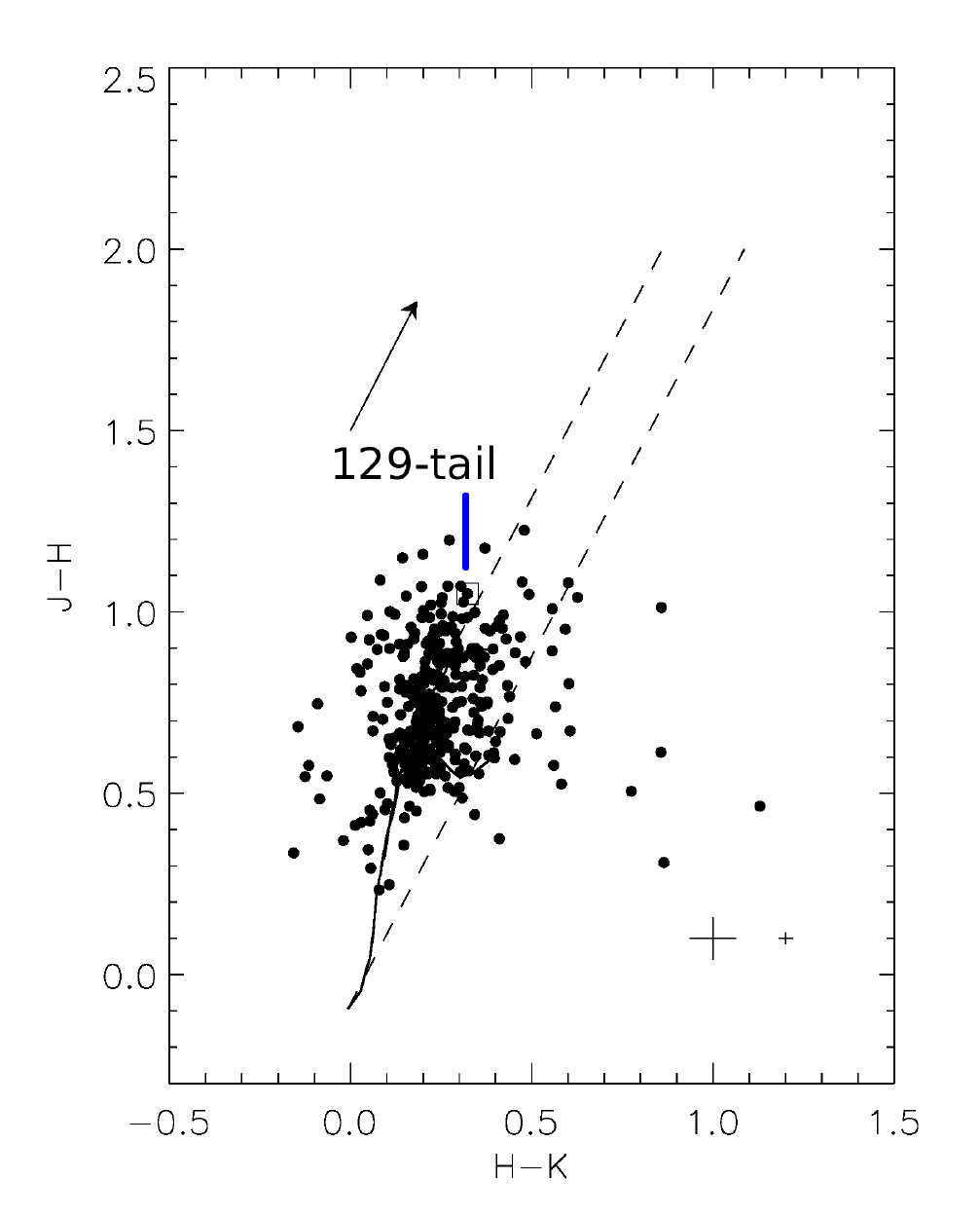}
    \caption{The \J$-$\H, \H$-$Ks\ color-color diagram of all objects in F19. Markings are as in Fig. \ref{fig:jhhkdiagramf7}. There is no continuous shell detected in F19, so all stars are counted as background stars and plotted in the same diagram. 129-tail marks a star seen in the more diffuse part of the tail in RN 129 and has been marked for visual reference.}
 \label{fig:jhhkdiagramf19}
 \end{figure*}

\begin{figure*}
 \centering
 \includegraphics [width=14cm]{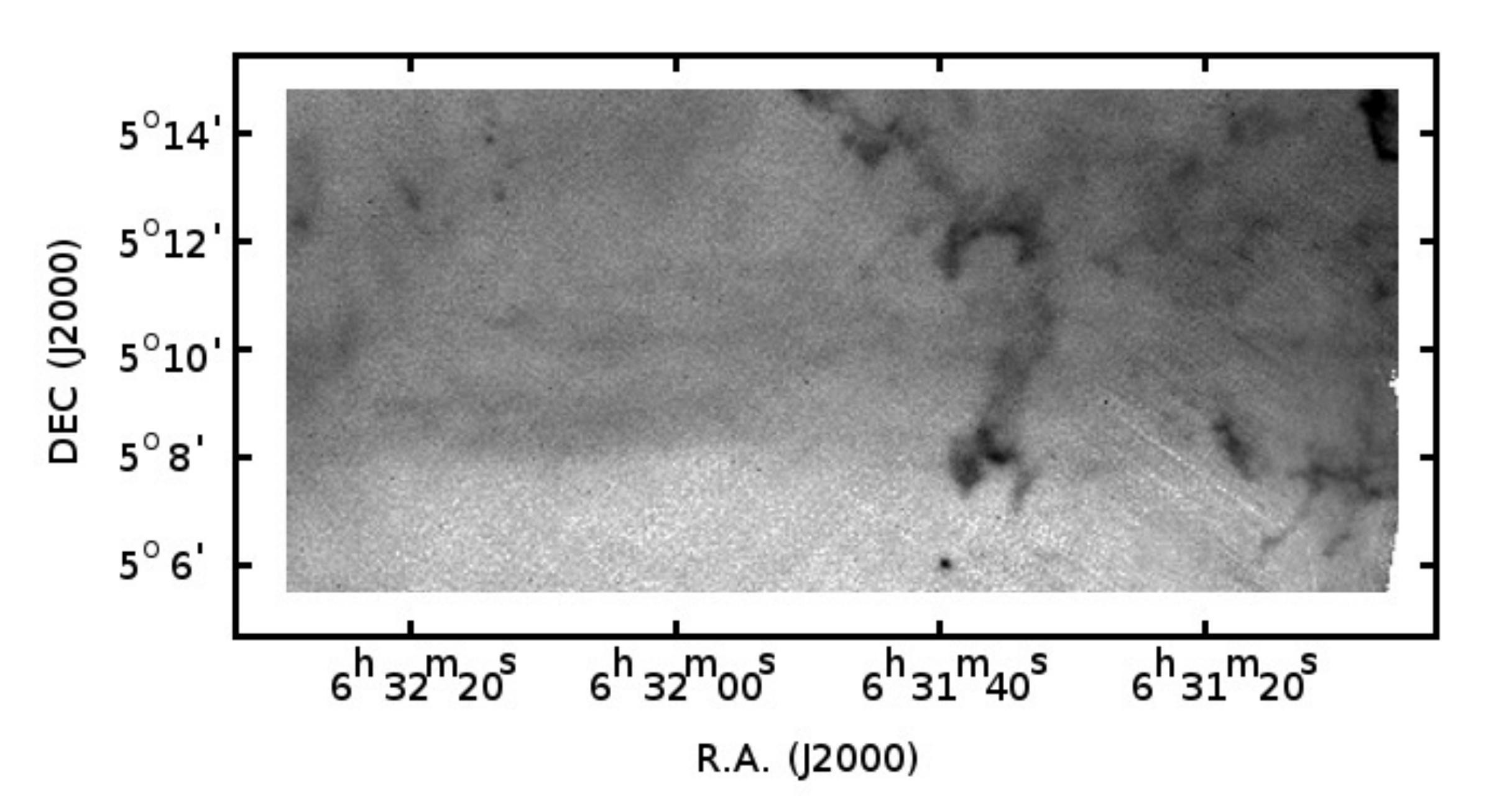}
    \caption{PACS 70 $\mum$ negative image covering fields 8, 14, 15, and 19.}
 \label{fig:pacs70neg}
 \end{figure*}

\begin{figure*}
 \centering
 \includegraphics [width=14cm]{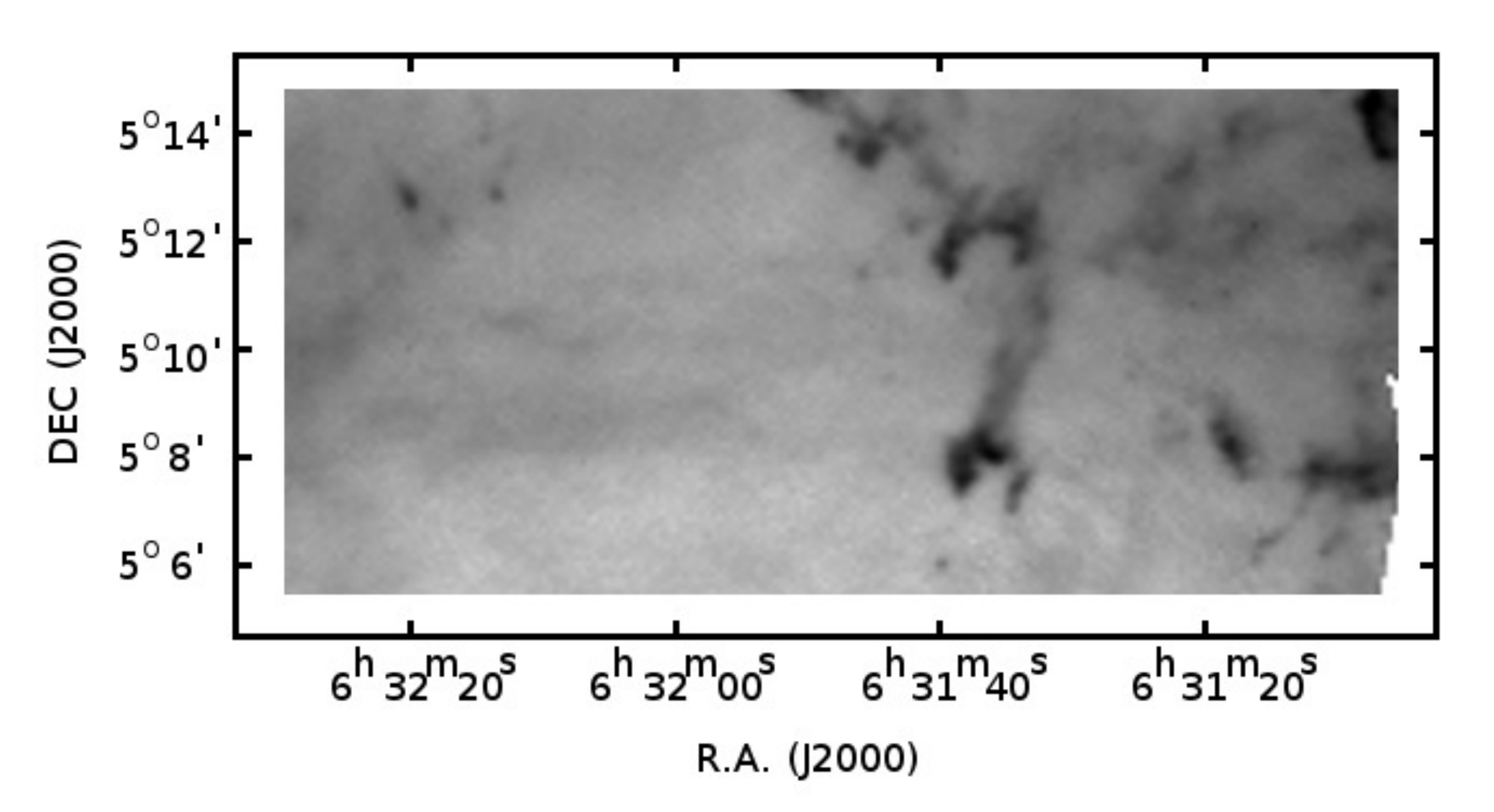}
    \caption{PACS 160 $\mum$ negative image covering fields 8, 14, 15, and 19.}
 \label{fig:pacs160neg}
 \end{figure*}

\end{appendix}

\end{document}